\documentclass[12pt, portrait]{emulateapj}
\usepackage{natbib}
\usepackage{mathrsfs}
\usepackage{url}

\newcommand{\kms}{\ifmmode {\rm km~s}^{-1} \else km~s$^{-1}$\fi}
\newcommand{\ergs}{\ifmmode {\rm erg~ s}^{-1} \else erg~s$^{-1}$\fi}
\newcommand{\ergscm}{\ifmmode {\rm erg~s}^{-1} \else erg~s$^{-1}$ cm$^{-2}$\fi}
\newcommand{\Msun}{\ifmmode {\rm M}_{\odot} \else M$_{\odot}$\fi }
\newcommand{\Lsun}{\ifmmode {\rm L}_{\odot} \else L$_{\odot}$\fi}
\newcommand{\qo}{\ifmmode q_{\rm o} \else $q_{\rm o}$\fi}
\newcommand{\Ho}{\ifmmode H_{\rm o} \else $H_{\rm o}$\fi}
\newcommand{\ho}{\ifmmode h_{\rm o} \else $h_{\rm o}$\fi}

\newcommand{\vFWHM}{\ifmmode v_{\mbox{\tiny FWHM}} \else
                    $v_{\mbox{\tiny FWHM}}$\fi}
\newcommand{\CCF}{\ifmmode F_{\it CCF} \else $F_{\it CCF}$\fi}
\newcommand{\ACF}{\ifmmode F_{\it ACF} \else $F_{\it ACF}$\fi}
\newcommand{\Halpha}{\ifmmode {\rm H}\alpha \else H$\alpha$\fi}
\newcommand{\Hbeta}{\ifmmode {\rm H}\beta \else H$\beta$\fi}
\newcommand{\Hgamma}{\ifmmode {\rm H}\gamma \else H$\gamma$\fi}
\newcommand{\Hdelta}{\ifmmode {\rm H}\delta \else H$\delta$\fi}
\newcommand{\Lya}{\ifmmode {\rm Ly}\alpha \else Ly$\alpha$\fi}
\newcommand{\Lyb}{\ifmmode {\rm Ly}\beta \else Ly$\beta$\fi}
\newcommand{\HeI}{\ifmmode {\rm He}\,{\sc i}\,\lambda5876 \else 
	          He\,{\sc i}\,$\lambda5876$\fi}
\newcommand{\HeII}{\ifmmode {\rm He}\,{\sc ii}\,\lambda4686 \else 
	           He\,{\sc ii}\,$\lambda4686$\fi}

\newcommand{\heii}{He\,{\sc ii}}

\newcommand{\ciii}{\ifmmode {\rm C}\,{\sc iii} \else C\,{\sc iii}\fi}
\newcommand{\civ}{C\,{\sc iv}}

\newcommand{\OIII}{[O\,{\sc iii}]$\lambda 5007$}

\newcommand{\mbh}{$M_{\rm BH}$\ }

\newcommand{\msigma}{$M_{\rm BH}$--$\sigma_{*}$\ }

\newcommand{\radlum}{$R_{\rm BLR}$--$L$\ }

\newcommand{\transfer}{$\Psi(V,\tau)$}

\def\fake2{\hphantom{3}}

\shorttitle{Velocity-Delay Maps}
\shortauthors{Grier et al.}

\begin{document}

\title{The Structure of the Broad Line Region in AGN: \\
I. Reconstructed Velocity-Delay Maps}

\author{C.~J.~Grier\altaffilmark{1},
B.~M.~Peterson\altaffilmark{1,2},
Keith~Horne\altaffilmark{3},
M.~C.~Bentz\altaffilmark{4},
R.~W.~Pogge\altaffilmark{1,2},
K.~D.~Denney\altaffilmark{5},
G.~De~Rosa\altaffilmark{1},
Paul~Martini\altaffilmark{1,2},
C.~S.~Kochanek\altaffilmark{1,2},
Y.~Zu\altaffilmark{1},
B.~Shappee\altaffilmark{1},
R.~Siverd\altaffilmark{6},
T.~G.~Beatty\altaffilmark{1},
S.~G.~Sergeev\altaffilmark{7},
S.~Kaspi\altaffilmark{8,9},
C.~Araya~Salvo\altaffilmark{1},
J.~C.~Bird\altaffilmark{1},
D.~J.~Bord\altaffilmark{10},
G.~A.~Borman\altaffilmark{7,11},
X.~Che\altaffilmark{12},
C.~Chen\altaffilmark{13},
S.~A.~Cohen\altaffilmark{13},
M.~Dietrich\altaffilmark{1},
V.~T.~Doroshenko\altaffilmark{7,14},
Yu.~S.~Efimov\altaffilmark{7,*},
N.~Free\altaffilmark{15},
I.~Ginsburg\altaffilmark{13},
C.~B.~Henderson\altaffilmark{1},
A.~L.~King\altaffilmark{12},
K.~Mogren\altaffilmark{1},
M.~Molina\altaffilmark{1,16},
A.~M.~Mosquera\altaffilmark{1},
S.~V.~Nazarov\altaffilmark{7,11},
D.~N.~Okhmat\altaffilmark{7,11},
O.~Pejcha\altaffilmark{1},
S.~Rafter\altaffilmark{9},
J.~C.~Shields\altaffilmark{15},
J.~Skowron\altaffilmark{1},
D.~M.~Szczygiel\altaffilmark{1},
M.~Valluri\altaffilmark{12},
and J.~L.~van~Saders\altaffilmark{1}
}

\altaffiltext{1}{Department of Astronomy, The Ohio State University,
140 W 18th Ave, Columbus, OH 43210} 
\altaffiltext{2}{Center for
Cosmology \& AstroParticle Physics, The Ohio State University, 191
West Woodruff Ave, Columbus, OH 4321, USA} 
\altaffiltext{3}{SUPA Physics \& Astronomy, University of St. Andrews, Fife, KY16 
9SS Scotland, UK}
\altaffiltext{4}{Department of Physics and Astronomy, Georgia State
University, Astronomy Offices, One Park Place South SE, Suite 700,
Atlanta, GA 30303, USA} 
\altaffiltext{5}{Marie Curie Fellow at the Dark
Cosmology Centre, Niels Bohr Institute, University of Copenhagen,
Juliane Maries Vej 30, DK-2100 Copenhagen, Denmark}
\altaffiltext{6}{Department of Physics and Astronomy, Vanderbilt University, 
5301 Stevenson Center, Nashville, TN 37235}
\altaffiltext{7}{Crimean Astrophysical
Observatory, P/O Nauchny Crimea 98409, Ukraine}
\altaffiltext{8}{School of Physics and Astronomy, Raymond and Beverly
Sackler Faculty of Exact Sciences, Tel Aviv University, Tel Aviv
69978, Israel} 
\altaffiltext{9}{Physics Department, Technion, Haifa
32000, Israel} 
\altaffiltext{10}{Department of Natural Sciences, The
University of Michigan~-~Dearborn, 4901 Evergreen Rd, Dearborn, MI
48128} 
\altaffiltext{11}{Isaac Newton Institute of Chile, Crimean
Branch, Ukraine} 
\altaffiltext{12}{Department of Astronomy, University
of Michigan, 500 Church Street, Ann Arbor, MI 41809}
\altaffiltext{13}{Department of Physics and Astronomy, Dartmouth
College, 6127 Wilder Laboratory, Hanover, NH 03755}
\altaffiltext{14}{South Station of the Moscow MV Lomonosov State University, 
Moscow, Russia, P/O Nauchny, 98409 Crimea, Ukraine} 
\altaffiltext{15}{Department of Physics \& Astronomy, Ohio
University, Athens, OH 45701} 
\altaffiltext{16}{Department of Astronomy \& Astrophysics, The Pennsylvania State 
University, 525 Davey Lab, University Park, PA 16802}
\altaffiltext{*}{Deceased, 2011 October 21}

\begin{abstract}
We present velocity-resolved reverberation results for five active
galactic nuclei. We recovered velocity-delay maps using the
maximum-entropy method for four objects: Mrk 335, Mrk 1501, 3C\,120,
and PG\,2130+099. For the fifth, Mrk~6, we were only able to measure
mean time delays in different velocity bins of the \Hbeta \ emission
line. The four velocity-delay maps show unique dynamical signatures
for each object. For 3C\,120, the Balmer lines show kinematic
signatures consistent with both an inclined disk and infalling gas,
but the \HeII \ emission line is suggestive only of inflow. The Balmer
lines in Mrk 335, Mrk 1501, and PG\,2130+099 show signs of infalling
gas, but the \heii \ emission in Mrk 335 is consistent with an
inclined disk. We also see tentative evidence of combined virial
motion and infalling gas from the velocity-binned analysis of Mrk
6. The maps for 3C\,120 and Mrk 335 are two of the most clearly
defined velocity-delay maps to date. These maps constitute a large
increase in the number of objects for which we have resolved
velocity-delay maps and provide evidence supporting the reliability of
reverberation-based black hole mass measurements.

\end{abstract}

\keywords{galaxies: active --- galaxies: nuclei --- galaxies: Seyfert}

\section{INTRODUCTION}
One of the continuing mysteries in the studies of active galactic
nuclei (AGN) is the structure and kinematics of the broad line region
(BLR). It is generally accepted that AGN are powered by supermassive
black holes, and the broad emission lines seen in Type 1 AGN are the
result of the photoionization of gas in the BLR. Under the assumption
that this gas is in virial motion around the black hole, we can use it
to measure the mass of the black hole itself. However, it is
impossible to study the structure of the BLR directly because the BLR
is on the order of light-days in radius, which renders it spatially
unresolvable even in the nearest of galaxies with the largest
diffraction-limited telescopes. The unknown BLR structure introduces
uncertainties in any black hole mass measurements derived from BLR gas
kinematics. To learn about the BLR structure, we must rely on either
reverberation mapping techniques (e.g., \citealt{Peterson04}) or
microlensing of gravitationally lensed quasars (e.g.,
\citealt{Guerras12}). Reverberation techniques use the time
variability observed in the AGN continuum emission and the subsequent
response of the gas in the BLR (\citealt{Blandford82};
\citealt{Peterson93}). By monitoring the AGN spectra over a period of
time, we can determine the distance of the emitting gas from the
central source by measuring the time delay between variations in the
continuum and the response of the emission lines. This time delay is
due to the light-travel time between the continuum source and the BLR.

The variations in the BLR emission-line flux, $\Delta L(V,t)$, are a
convolution of the continuum flux variations, $\Delta C(t)$, with the
``transfer function'', $\Psi(V,\tau)$ (\citealt{Blandford82}). The
transfer function depends on the temporal lag $\tau$ between the line
and continuum emission, and the line-of-sight velocity $V$ of the BLR
gas. The relationship is expressed mathematically as
\begin{equation}
\label{eq:trans}
\Delta L(V,t) = \int_{0}^{\infty} \Psi (V,\tau)\Delta C(t-\tau )d\tau.
\end{equation}
A main goal of reverberation mapping is to recover \transfer, called
the ``velocity-delay map'', which describes how the continuum flux
variations give rise to BLR flux variations, and therefore contains
information about the BLR geometry and kinematics. For example, a
Keplerian disk produces a velocity-symmetric structure with a
wider/narrower range of velocities at smaller/larger delays. In
contrast, radial flows give rise to asymmetric velocity structure, the
signature of infall/outflow being smaller delays on the red/blue side
of the velocity profile. Most previous reverberation studies have been
limited to measuring the mean time delay $\langle\tau\rangle$ for
various emission lines. This allows us to estimate the mean radius of
the BLR (e.g., \citealt{Peterson04}; \citealt{Denney10};
\citealt{Bentz10a}; \citealt{Grier12b}), but it reveals very little
information about the detailed structure of the BLR.

It has long been recognized that measuring emission-line time lags as
a function of line-of-sight velocity provides a way to determine the
gross kinematics of the BLR (e.g., \citealt{Bahcall72};
\citealt{Blandford82}; \citealt{Bochkarev82}; \citealt{Capriotti82};
\citealt{Antokhin83}), although it is only relatively recently that
suitable data have become available to achieve this in practice. There
were a number of early attempts to extract velocity-dependent time
lags in emission lines in order to probe the BLR kinematics, though
these were generally frustrated by low data quality, time-sampling
issues, or both. The first attempts that we are aware of to search for
velocity-dependent lags were by \cite{Gaskell88} for NGC\,4151 and
\cite{Koratkar89} for Fairall 9, in both cases for the C\,{\sc
iv}\,$\lambda1549$ emission line based on data obtained with the {\it
International Ultraviolet Explorer} (a mean lag for Fairall 9 had been
obtained earlier by \citealt{Clavel89}). In both cases, detection of
infall was claimed. We have, however, reanalyzed both of these data
sets using more modern methods and find that the light curves are far
too poorly sampled to support either claim of a detection of
infall. Later, \cite{Clavel90} revisited the case of NGC\,4151 with
improved time sampling. To within the accuracy of their measurement,
they found the red and blue wings of \civ \ to vary
simultaneously. However, their mean sampling interval of 3.4 days was
only slightly smaller than the \civ \ mean response time, which would
have made it difficult to detect the subtle velocity-dependent lags
that have been detected in optical lines in recent years (
\citealt{Denney09c}; \citealt{Bentz10a}).

The first well-sampled AGN emission-line light curves became available
as a result of intensive monitoring of NGC\,5548 with {\it IUE}
(\citealt{Clavel91}) and ground-based telescopes
(\citealt{Peterson91}; \citealt{Dietrich93}) in 1988--89. These data
led to the first successful attempts to recover one-dimensional delay
maps, $\Psi(\tau)$ (\citealt{Horne91}; \citealt{Krolik91}). Other
attempts were also made to search for velocity-dependent lags (e.g.,
\citealt{Clavel91b}, for NGC\,5548), but they generally concluded that
any velocity-dependent lags remained unresolved. However,
\cite{Crenshaw90} argued that C\,{\sc iv} profile variations during
part of the 1988--89 campaign on NGC\,5548 indicated infall of BLR
gas, although an infall signature was not detected using any other
subset of the campaign data and the {\it IUE} spectra were generally
rather marginal for AGN emission-line profile studies. Other sources
were searched for velocity-dependent emission-line responses, but
these searches were also unsuccessful (e.g.,
\citealt{Kollatschny97}). Some limited success was achieved with both
space-based observations (\citealt{Ulrich96}; \citealt{Wanders95}) and
ground-based data (\citealt{Kollatschny03}), but except in the case of
NGC\,4151 (\citealt{Ulrich96}) the resulting velocity-delay maps
showed little clear structure. There was also a growing understanding
that the dramatic emission-line profile variations observed in some
sources were not a reverberation effect (e.g., \citealt{Wanders96};
\citealt{Sergeev07}), and that reverberation signals were generally
going to be weak. Obtaining a high-fidelity velocity--delay map was
going to require both high-quality data and excellent temporal
sampling (\citealt{Horne04}). More recent spectroscopic monitoring
campaigns for reverberation-mapping have been specifically designed
to, among other things, recover velocity-delay maps (e.g.,
\citealt{Bentz10a}, \citealt{Denney11}). These efforts have already
been shown to be successful --- \cite{Bentz10b} show well-resolved
velocity-delay maps for Arp 151 using maximum-entropy methods, and
\cite{Pancoast12} used dynamical modeling (first described by
\citealt{Pancoast11}) to recover BLR structure information in Mrk 50.

In late 2010, we carried out a four-month long reverberation mapping
campaign with the ultimate goal of recovering velocity-delay maps for
several of the targets. Details of the data processing, light curves,
mean time lags, and black hole mass measurements for these objects
were published by \cite{Grier12b}. Here we present velocity-binned
reverberation results for all five objects observed in our 2010
campaign and two-dimensional velocity-delay maps for four objects. All
five targets and their coordinates and redshifts are listed in Table
\ref{Table:obj_info}. While our reverberation campaign was aimed
primarily at investigating the \Hbeta \ emission line, in a few cases
we recover velocity-delay maps for the H$\gamma$ and \HeII \ emission
lines as well.

\section{DATA}
\label{sec:data}
For this study we use spectroscopic and photometric data obtained
during our reverberation campaign carried out at multiple institutions
in late 2010. The primary set of spectra were obtained using the
Boller and Chivens CCD spectrograph on the MDM 1.3m McGraw-Hill
telescope on Kitt Peak. We supplemented our spectroscopic continuum
light curves with photometry from the 46-cm Centurion telescope at
Wise Observatory of Tel-Aviv University and the 70-cm telescope at the
Crimean Astrophysical Observatory (CrAO), and with spectra obtained
from the 2.6m Shajn telescope at CrAO. All data were obtained between
2010 August 21 and 2011 January 7. The data processing is described in
detail in \cite{Grier12b}. In short, we calibrated the reduced spectra
onto an absolute flux scale by assuming that the flux in the narrow
\OIII \ emission line is constant. We created a reference spectrum for
each object using the spectra taken on photometric nights during the
campaign. We then used the procedure of \cite{vanGroningen92} to apply
small wavelength shifts and scale each individual spectrum to match
the flux in the \OIII \ emission line in the reference spectrum using
a $\chi^2$ goodness-of fit estimator method. The resulting scaled
spectra were used to create the driving continuum light curves used
throughout our analysis as well as the light curves used in our
velocity-binned time series analysis.

%
\section{INITIAL H{\boldmath $\beta$} \ VELOCITY-BINNED TIME SERIES ANALYSIS}
\label{sec:velres}
As a preliminary step, we first searched for gross kinematic
signatures by seeing if different parts of each emission line show
different time delays with respect to the continuum. Following
\cite{Denney09c}, we divided the \Hbeta \ emission line into velocity
bins as follows: First, we divide the \Hbeta \ emission lines in half
at the zero-velocity line center (the systemic redshift), to separate
red-shifted and blue-shifted signals. We then divide each line half
into bins containing equal flux in the rms residual spectrum, choosing
the number of bins for each object based on the width of the rms line
profiles. We create light curves for individual bins by integrating
the flux within each bin in each scaled spectrum. The resulting light
curves were analyzed using JAVELIN, the updated version of SPEAR (see
\citealt{Zu11} and \citealt{Grier12b} for details), to measure the
time delay in each light curve with respect to the continuum light
curve. In essence, JAVELIN uses a statistical model of the continuum
light curve and its covariances and simultaneously fits the continuum
and line light curves assuming a simple top-hat transfer function.
This leads to a statistically well-defined means of interpolating the
irregularly sampled line and continuum data that essentially averages
over all possible interpolated light curves weighted by their
likelihood of fitting the data. In particular, the model fills gaps in
the light curves in a well-defined manner with well-defined
uncertainties.

\begin{figure}
\begin{center}
\epsscale{0.43}
\plotone{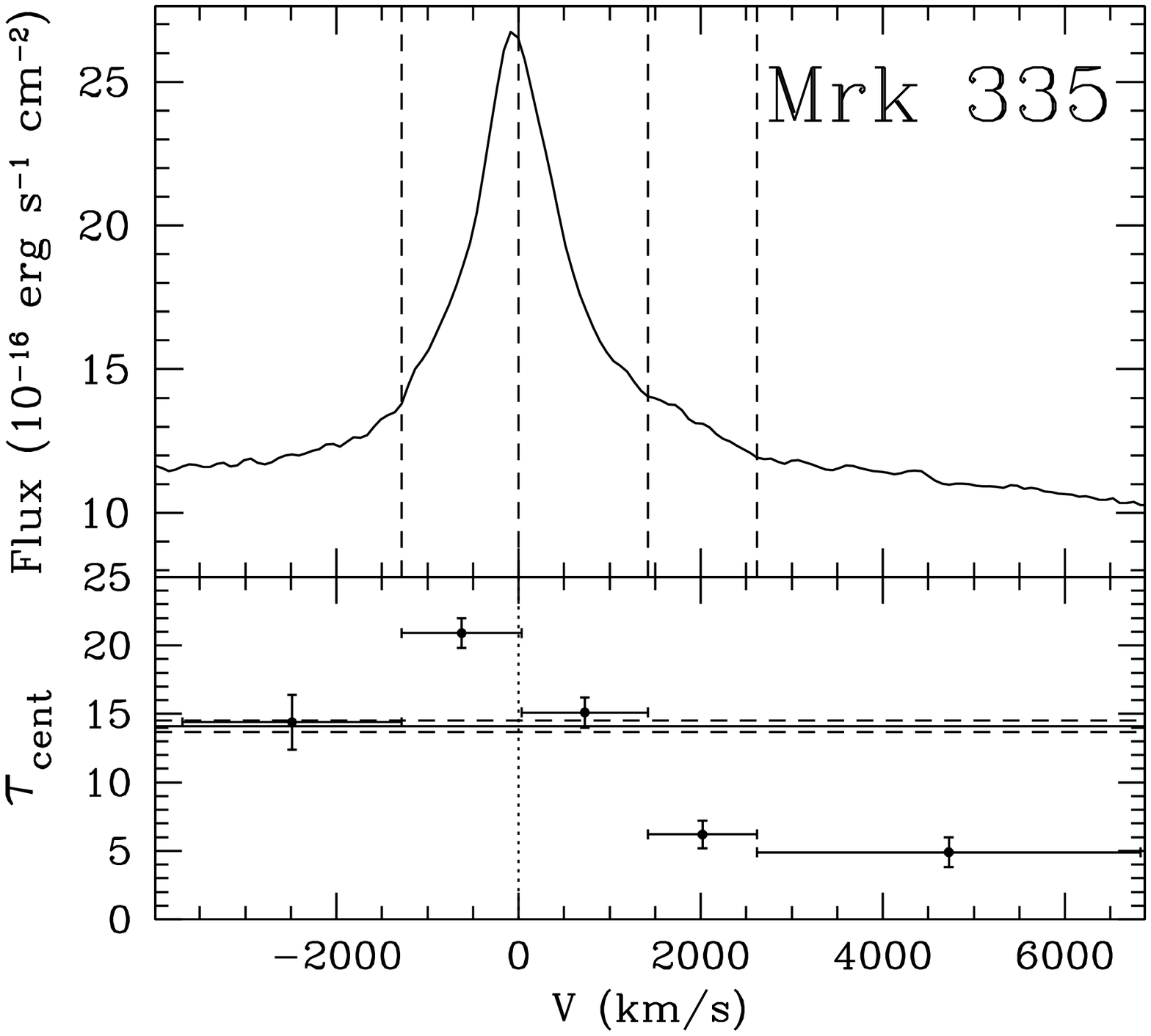}
\plotone{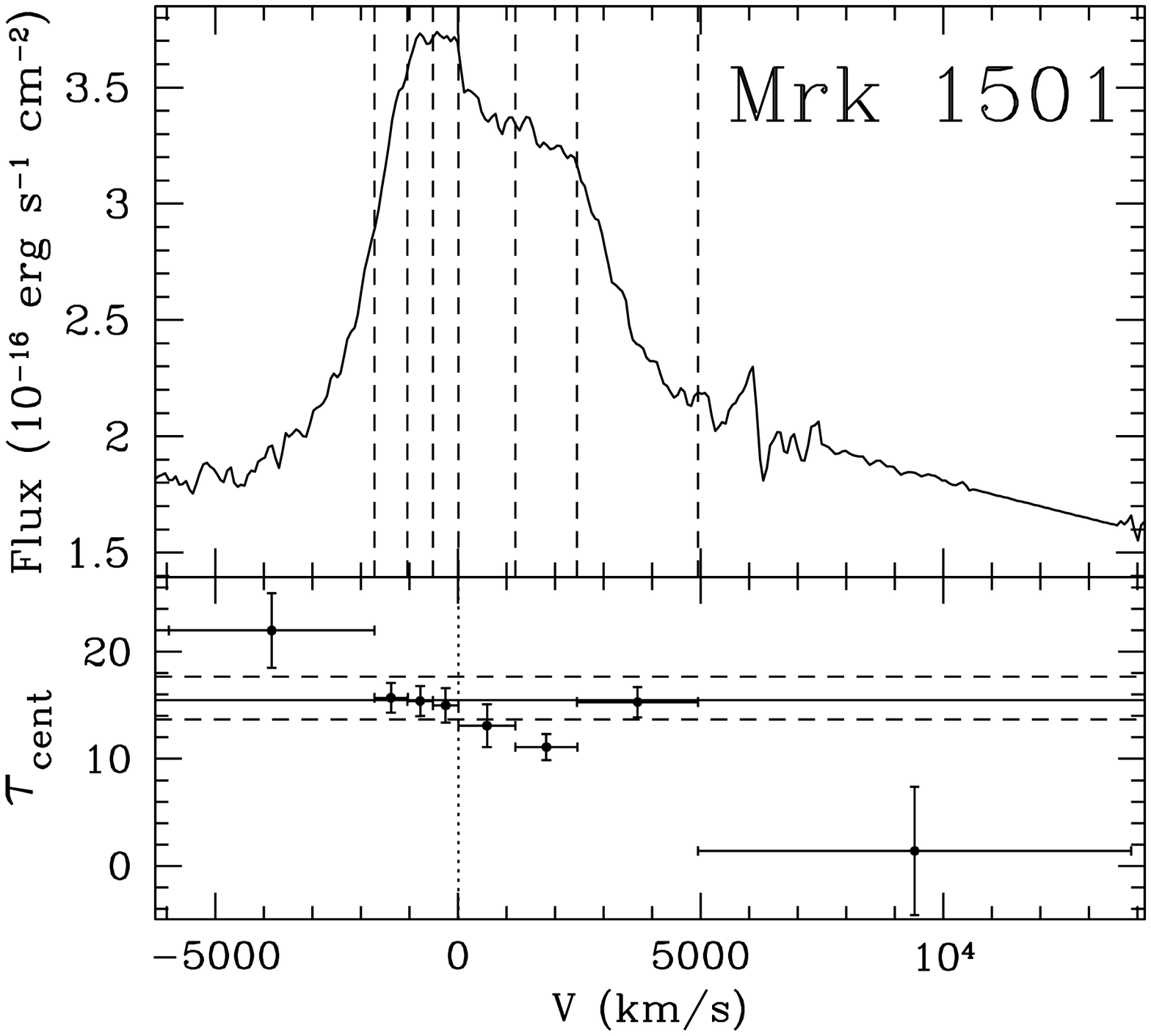}
\plotone{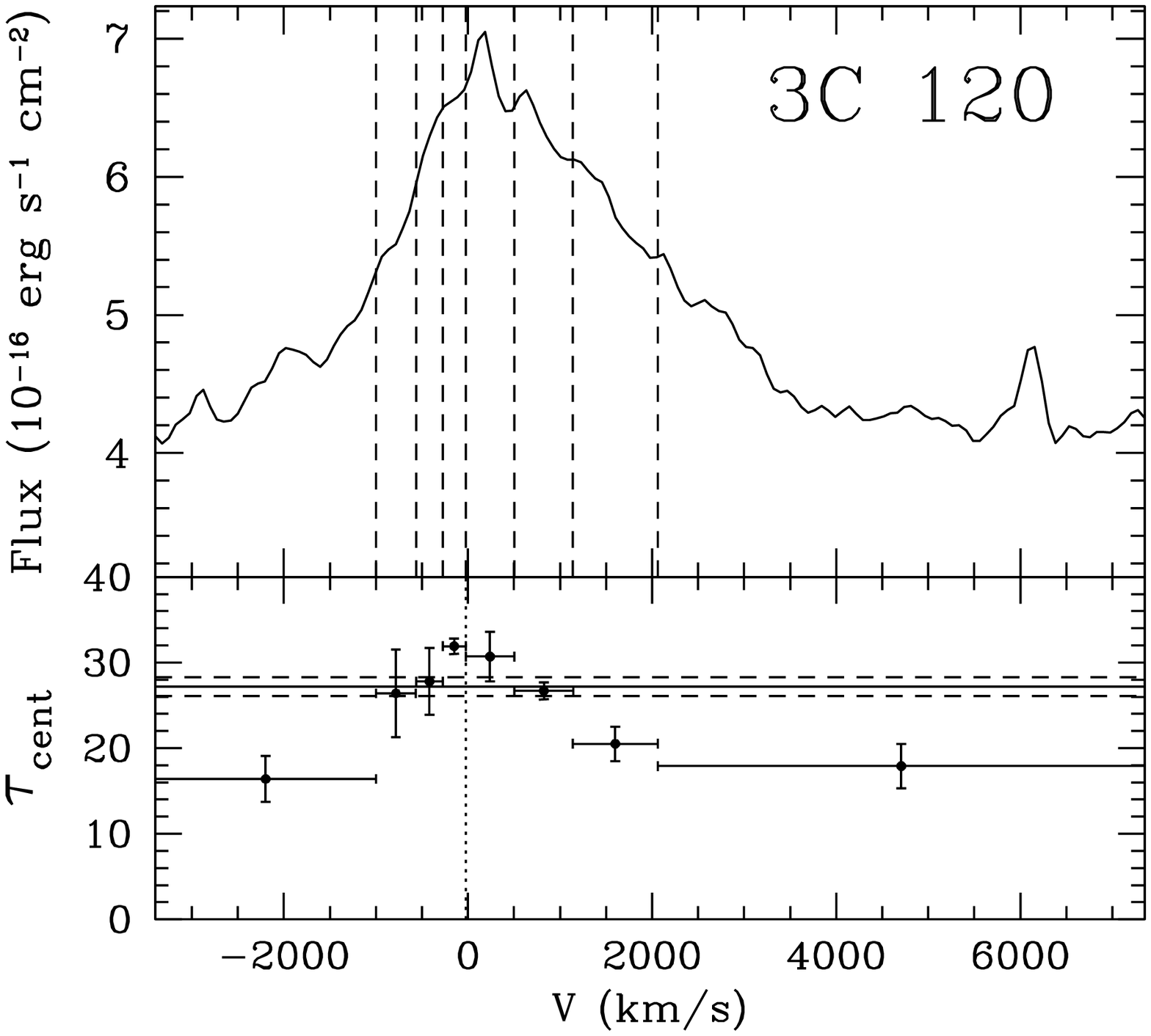}
\plotone{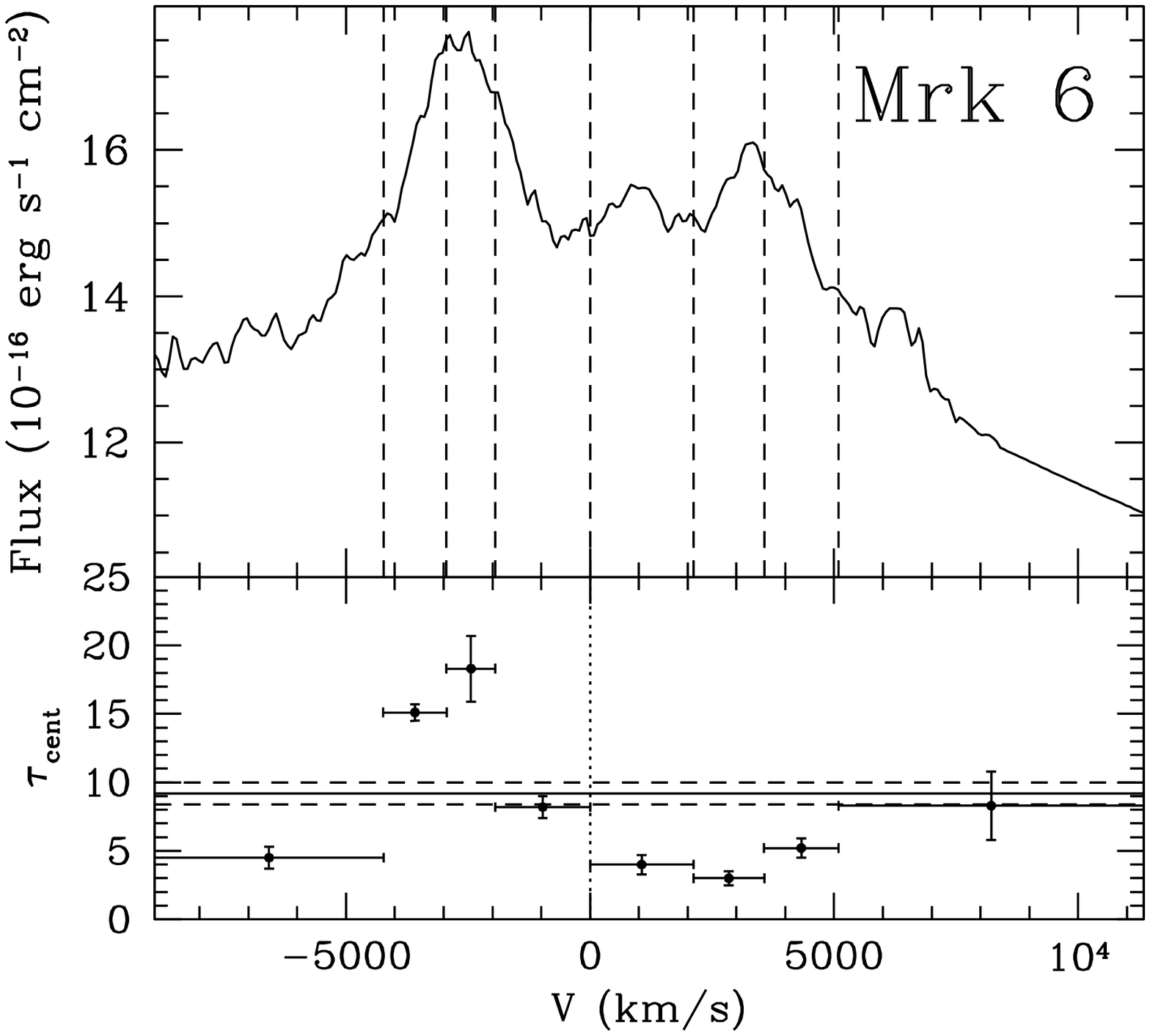}
\plotone{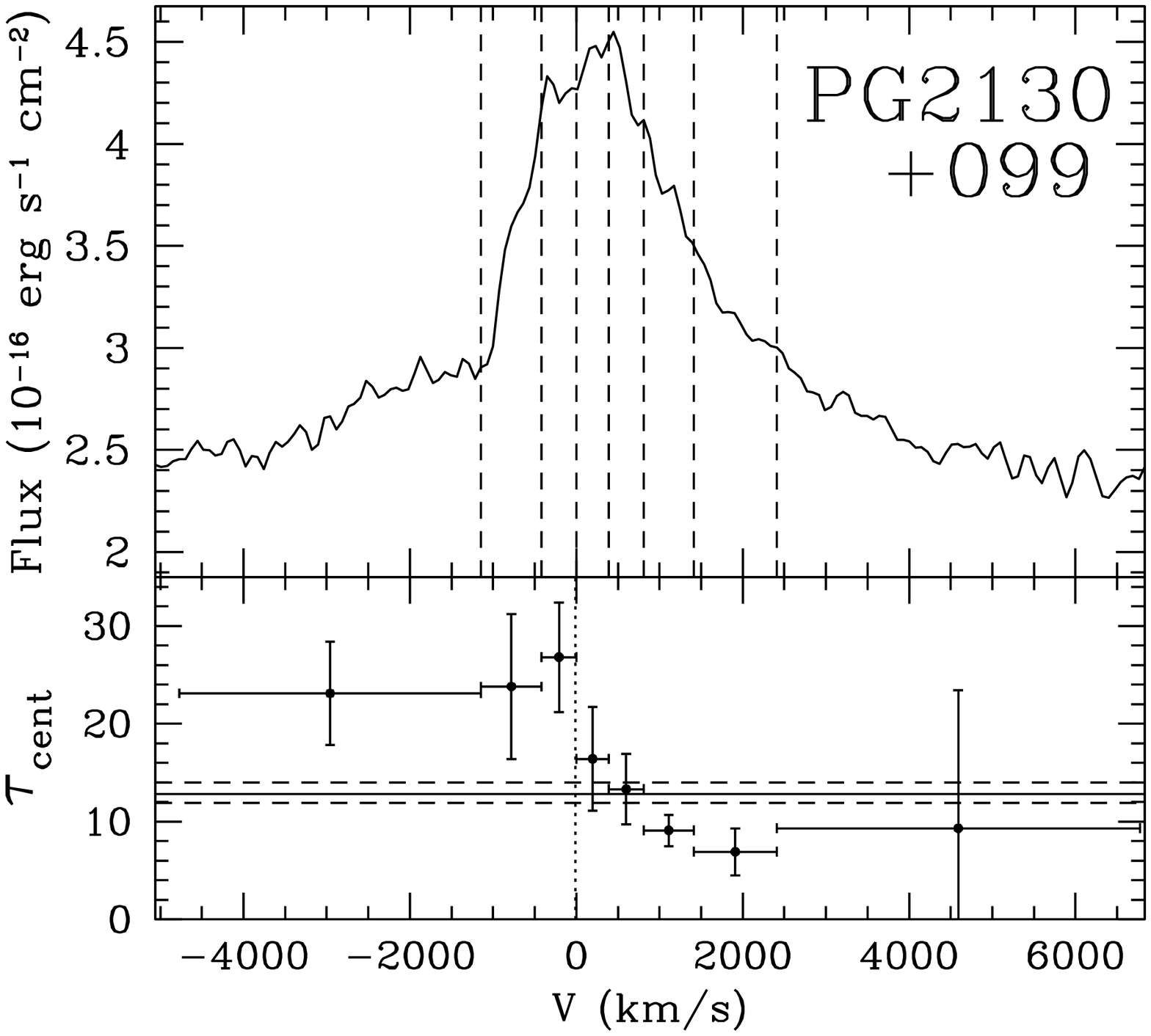}
\caption{Velocity-binned reverberation lag results. The top panels
show the rms residual spectrum for each object, with the edges of the
bins designated by vertical dashed lines. The bottom panels show the
mean time delays measured for each bin. The zero-velocity center of
the \Hbeta \ emission line is shown by the dotted line. The horizontal
solid line shows the average time lag reported in \cite{Grier12b},
with uncertainties as horizontal dashed lines. Error bars in the
velocity direction show the width of the bins.}
\label{fig:velres}
\end{center}
\end{figure}

Our results from JAVELIN are shown in Figure \ref{fig:velres}. The top
panels show the rms residual line profile and the wavelength bins, and
the bottom panel shows the distribution of lags for each part of the
emission line. As a check, we also ran the light curves for each bin
through a cross correlation analysis routine that is widely used in
reverberation mapping (e.g., \citealt{Peterson04}). The lags from the
cross correlation were nearly identical to the ones obtained from
JAVELIN. We show here only the JAVELIN results, since the JAVELIN
uncertainties are much smaller than those obtained through cross
correlation methods and the overall signatures are the same. In all
five objects, we see velocity-dependent time lags in the \Hbeta \
emission line. Mrk 335, Mrk 1501, and PG\,2130+099 all show longer
time lags on the blue side of the line profile than on the red side,
which is a signature of inflowing gas. By contrast, 3C\,120 shows a
nearly symmetric profile, with small lags in the outer wings, and
larger lags towards the line center, which is suggestive of a
disk. Mrk 6 shows more complex structures that could be a combination
of a disk plus infall, as previously suggested by \cite{Doroshenko12}
for this object.

\section{TWO-DIMENSIONAL MEMECHO FITTING}
\label{sec:memecho}
\subsection{Preparation of Spectra}
To recover the velocity-delay maps, we used maximum entropy methods as
implemented in the program MEMECHO (see \citealt{Horne91} and
\citealt{Horne94} for details). To prepare the data for use with
MEMECHO, we used software developed by Keith Horne called
PREPSPEC. PREPSPEC applies corrections to account for wavelength
shifts due to instrument flexure and differential refraction, spectrul
blurring due to seeing and instrumental resolution, and errors due to
slit losses and changes in atmospheric transmission between epochs. To
do this, we fit a calibration model to the spectra that accounts for
both the spectral variability and the abovementioned systematic
errors. We model the spectra as the sum of five different components:
1) a constant ``mean'' spectrum, which accounts for non-varying
components of the continuum and broad emission lines as well as
starlight from the host galaxy, 2) non-varying narrow emission lines,
3) a time-variable continuum, 4) time variable broad emission lines,
and 5) factors to account for wavelength shift, spectral blurring, and
scaling. After subtracting the models from the spectra, the resulting
continuum-subtracted line profile variations are used as inputs to
MEMECHO.

\subsection{MEM Fitting}
In brief, MEMECHO finds the
``simplest'' linearized echo model that fits the observed continuum
and emission-line spectral variations. This is accomplished by
minimizing the function
\begin{equation}
\label{eq:q}
Q=\chi^2-2\,\alpha\,S
\end{equation}
for a given model fit to a set of data. Here $\chi^2$ measures the
``badness-of-fit'' between $N$ data points with values $D_k$ and the
corresponding model predictions $\mu_k$ for those values, assuming
Gaussian errors with known variances $\sigma_k^2$. The entropy $S$
measures the ``simplicity'' of the model, elaborated below, and the
regularization parameter $\alpha$ controls the trade-off between these
competing requirements. MEMECHO adjusts $\alpha$ and the model
parameters $p_i$ to achieve a user-specified value of $\chi^2/N$ while
maximizing $S$.

In MEMECHO's linearized echo model 
\begin{equation}
L(\lambda,t) = L_0(\lambda) + \int_{0}^{\infty} \Psi(\lambda,\tau)(C(t-\tau)-C_0)d\tau,
\end{equation}
the parameters $p_i$ include the continuum light curve $C(t)$ on an
evenly spaced grid, the delay map at each wavelength
$\Psi(\lambda,\tau)$, and the time-independent background spectrum
$L_0(\lambda)$. $C_0$ is a reference continuum level, which we set at
the median of the continuum light curve data. The entropy of the model
is defined as
\begin{equation}
\label{eq:entropy}
        S = \sum_i w_i( p_i - q_i - p_i \ln{(p_i/q_i)}),
\end{equation}
where $w_i$ are weights, $p_i$ are the positive parameters outlined
above, and $q_i$ are the default values of these parameters.  Note
that $S$ is maximized when $p_i=q_i$. Minimizing $Q$ gives
\begin{equation}
        0 = \sum_{k=1}^{N} \frac{D_{k}-\mu_{k}}{\sigma^2_k} \frac{d\mu_k}{dp_i}
        + \alpha \, w_i \, \ln{(p_i/q_i)}.
\end{equation}
Thus the model parameters $p_i$ are pulled by the data toward
$\mu_k=D_k$ and by the entropy toward $p_i=q_i$, with $\alpha$
adjusting the trade-off between the two.  With default
values $q_i=(p_{i-1}\,p_{i+1})^{1/2}$, the entropy ``pulls'' each
$p_i$ toward the geometric mean of its neighbors, so that the entropy
penalizes regions of high curvature and favors smooth functions with
gaussian features and exponential tails. The weights $w_i$ and default
values $q_i$ are assigned with two parameters in MEMECHO: $W$ to stiffen
$\Psi(\lambda,\tau)$ relative to $C(t)$, and $A$ to control the aspect
ratio of features in $\Psi(\lambda,\tau)$. The $A$ and $W$ parameters
are set by the user. Their qualitative effects on the velocity-delay
maps are described below in Section \ref{sec:aandw}.

\subsection{The Driving Continuum Light Curve}
\begin{figure}
\begin{center}
\epsscale{1.0}
\plotone{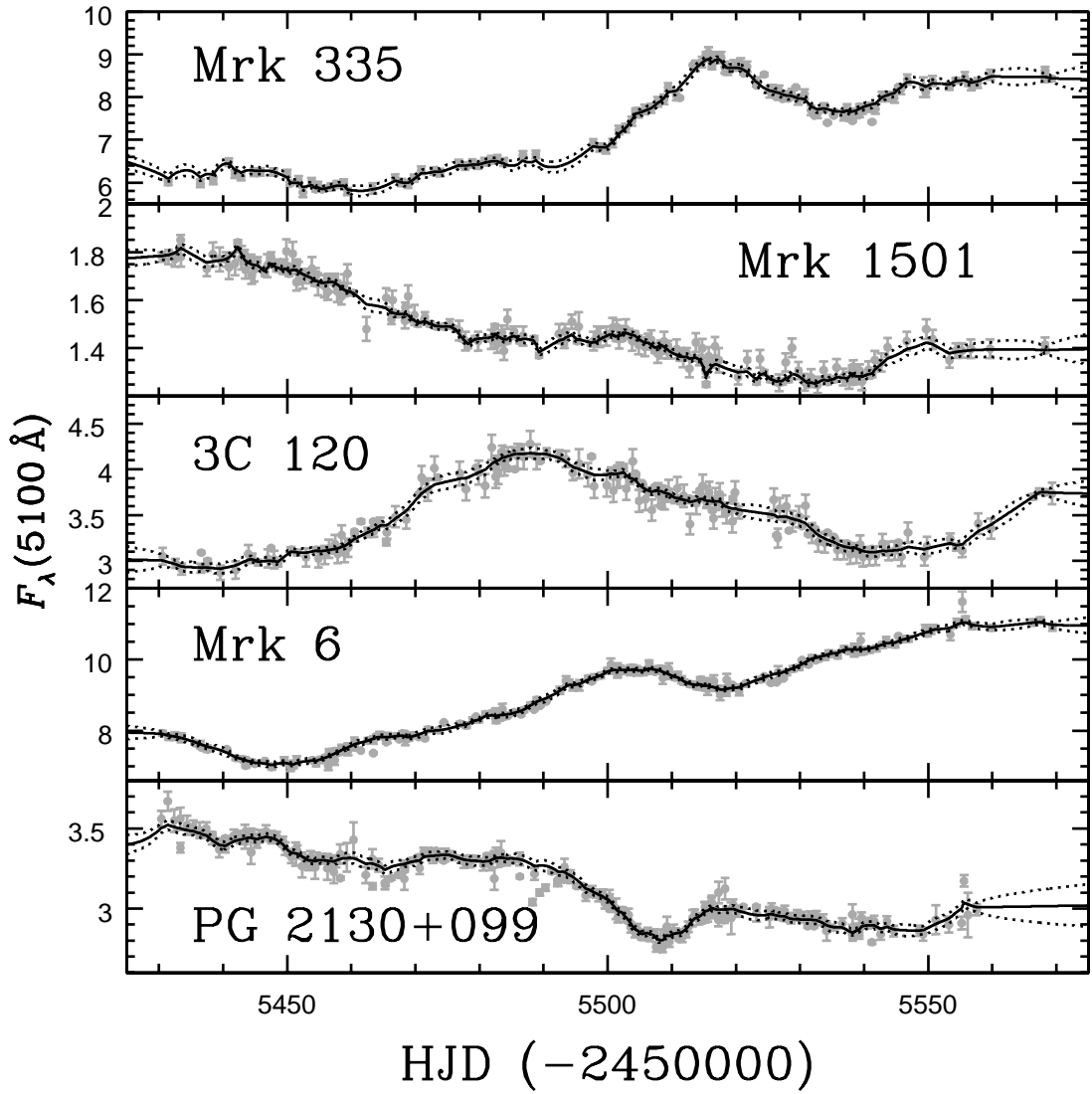}
\caption{Original continuum light curves (gray data points) from
\cite{Grier12b} and the JAVELIN mean continuum model (black solid
line) used in the MEMECHO analysis. The dotted line shows the standard
deviation about the mean light curve from individual model
realizations. Fluxes are given in units of 10$^{-15}$ \ergscm
\AA$^{-1}$.}
\label{fig:javmodels}
\end{center}
\end{figure}

One of the major practical issues with the use of MEMECHO is its
tendency to introduce spurious features at gaps in the continuum light
curve to drive $\chi^2/N$ to the minimum value set by the user. This
becomes a significant problem as the target $\chi^2/N$
decreases. Altering MEMECHO parameters to stiffen (i.e., penalize
rapid variations in) the driving light curve is sometimes helpful, but
we were still unable to produce velocity-delay maps with any
discernible structure using the original continuum light curves. To
provide stronger constraints on the driving light curve model, we ran
our full continuum light curves (including both spectroscopic and
photometric data) through the JAVELIN modeling software of
\cite{Zu11}. This method models the continuum with a damped random
walk (DRW) model that has been demonstrated to be a good statistical
model of AGN variability (e.g., \citealt{Kelly09};
\citealt{Kozlowski10}; \citealt{MacLeod10}; \citealt{MacLeod12};
\citealt{Zu12}). This allowed us to create highly sampled {\it
simulated} continuum light curves with reliable uncertainties that
represent the range of the most likely continuum behavior within the
gaps. Instead of providing MEMECHO with our original continuum light
curve, we use our highly sampled JAVELIN mean light curve, which more
strongly constrains the MEMECHO continuum model. This light curve is
the probability-weighted mean of DRW light curves consistent with the
data and the DRW model. Uncertainties on each point in the simulated
light curve represent the standard deviation of probable light curves
around this mean. These narrow to match the measurement errors of the
data points and then broaden as the gaps between the data become
larger. The variability observed in the adopted mean simulated light
curve is somewhat smoothed compared to an individual DRW model
realization, but the error envelope applied to the mean light curve,
representing the 1$\sigma$ deviations of the individual realizations
about the mean, accounts for these differences. The original continuum
light curves and their corresponding simulated light curves are shown
in Figure~\ref{fig:javmodels}.

\begin{figure}
\begin{center}
\epsscale{0.7}
\plotone{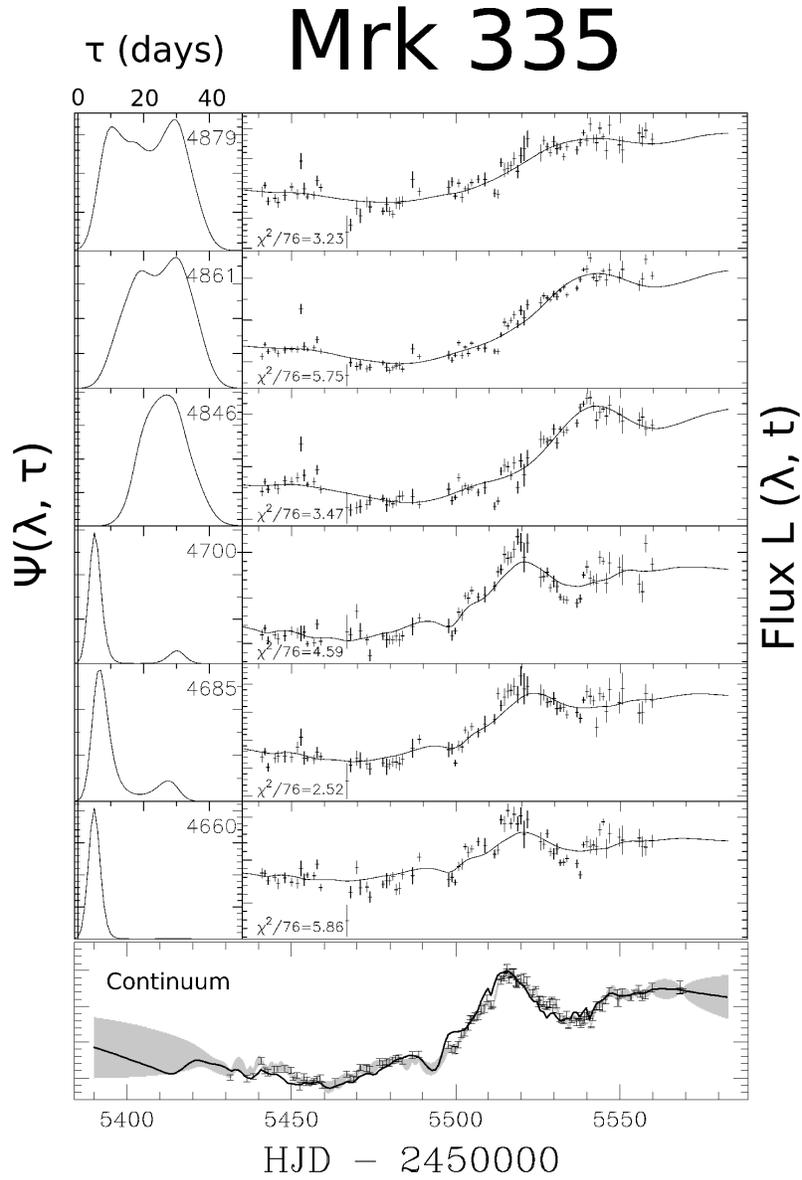}
\caption{Best MEMECHO fits to the spectra of Mrk 335. The left panels
show the delay maps at each selected rest-frame wavelength, given in
the top righthand corner of each panel. The right panels show the
emission-line light curve at these selected wavelengths and the
MEMECHO fit to the light curve. The $\chi^2/N$ of the fit for each
light curve is shown in each panel. The bottom panel shows the
original continuum light curve (black error bars), the error envelope
from the simulated light curve showing the standard deviation about
the mean (gray envelope), and the continuum model from MEMECHO (solid
black line).}
\label{fig:mrk335fits}
\end{center}
\end{figure}

\begin{figure}
\begin{center}
\epsscale{0.8}
\plotone{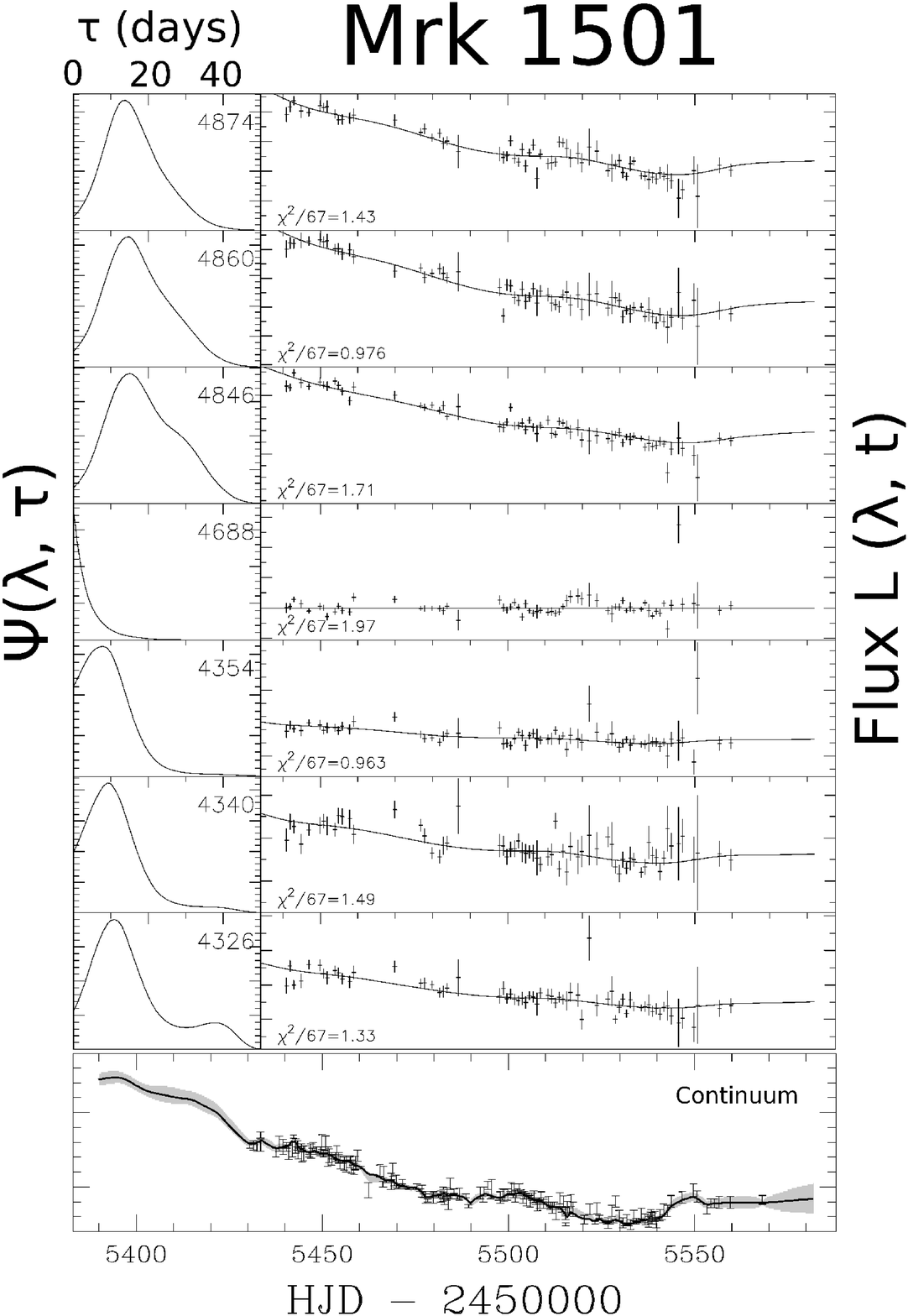}
\caption{Best MEMECHO fits to the spectra of Mrk 1501. Panels and
symbols are the same as in Figure \ref{fig:mrk335fits}.}
\label{fig:mrk1501fits}
\end{center}
\end{figure}

\begin{figure}
\begin{center}
\epsscale{0.8}
\plotone{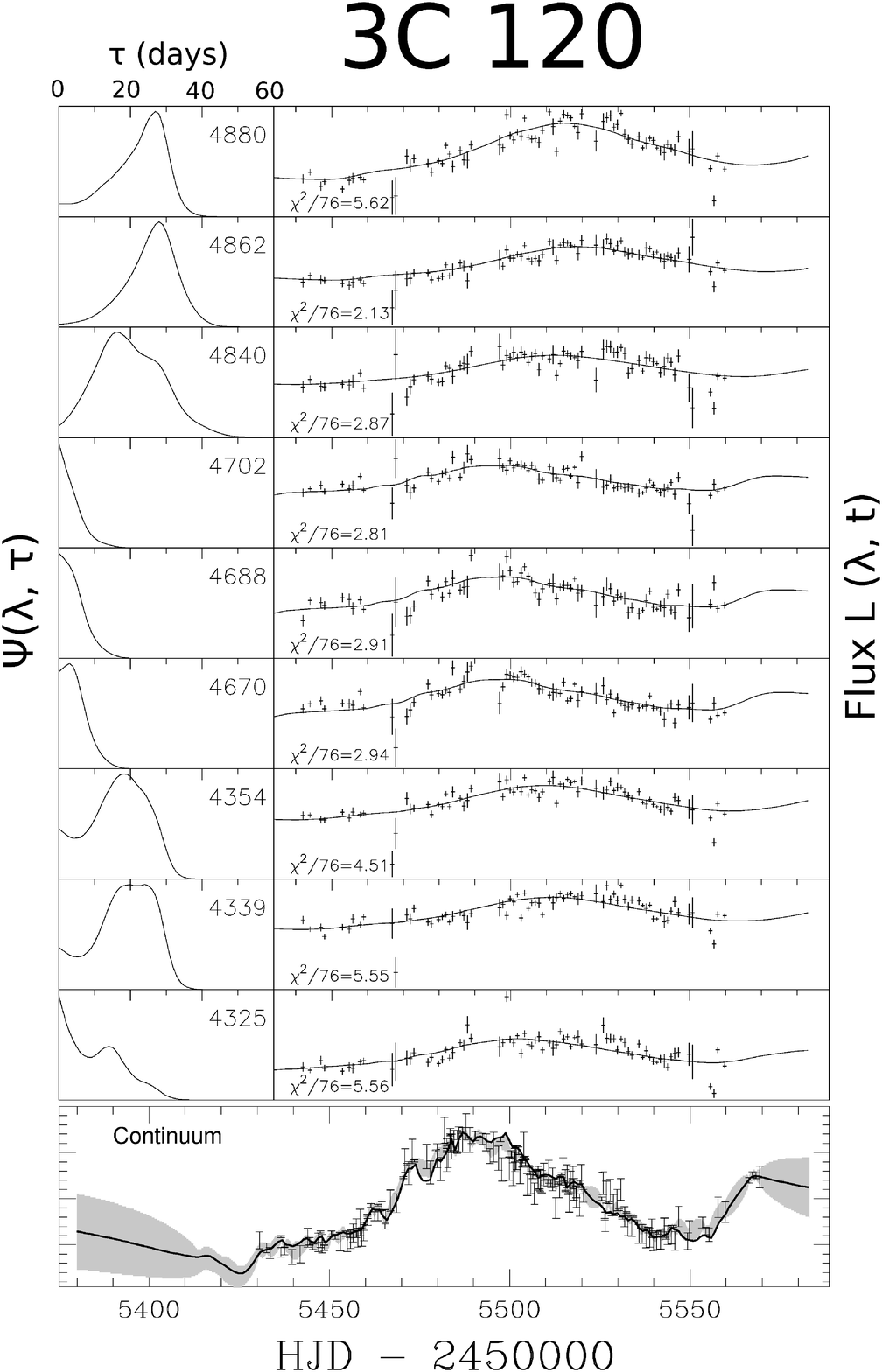}
\caption{Best MEMECHO fits to the spectra of 3C\,120. Panels and
symbols are the same as in Figure \ref{fig:mrk335fits}. }
\label{fig:3c120fits}
\end{center}
\end{figure}

\begin{figure}
\begin{center}
\epsscale{0.8}
\plotone{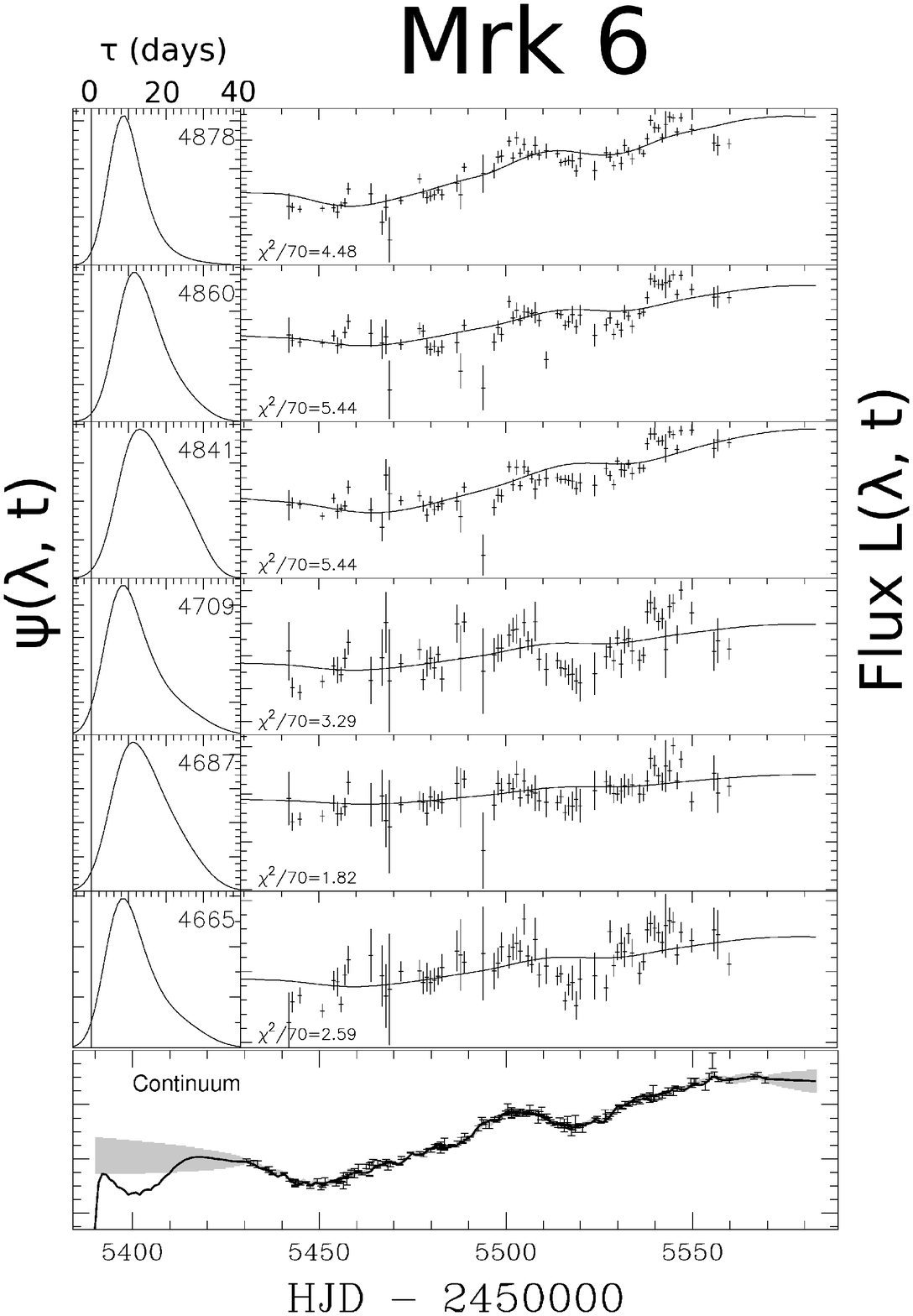}
\caption{Best MEMECHO fits to the spectra of Mrk 6. Panels and symbols
are the same as in Figure \ref{fig:mrk335fits}. }
\label{fig:mrk6fits}
\end{center}
\end{figure}

\begin{figure}
\begin{center}
\epsscale{0.8} \plotone{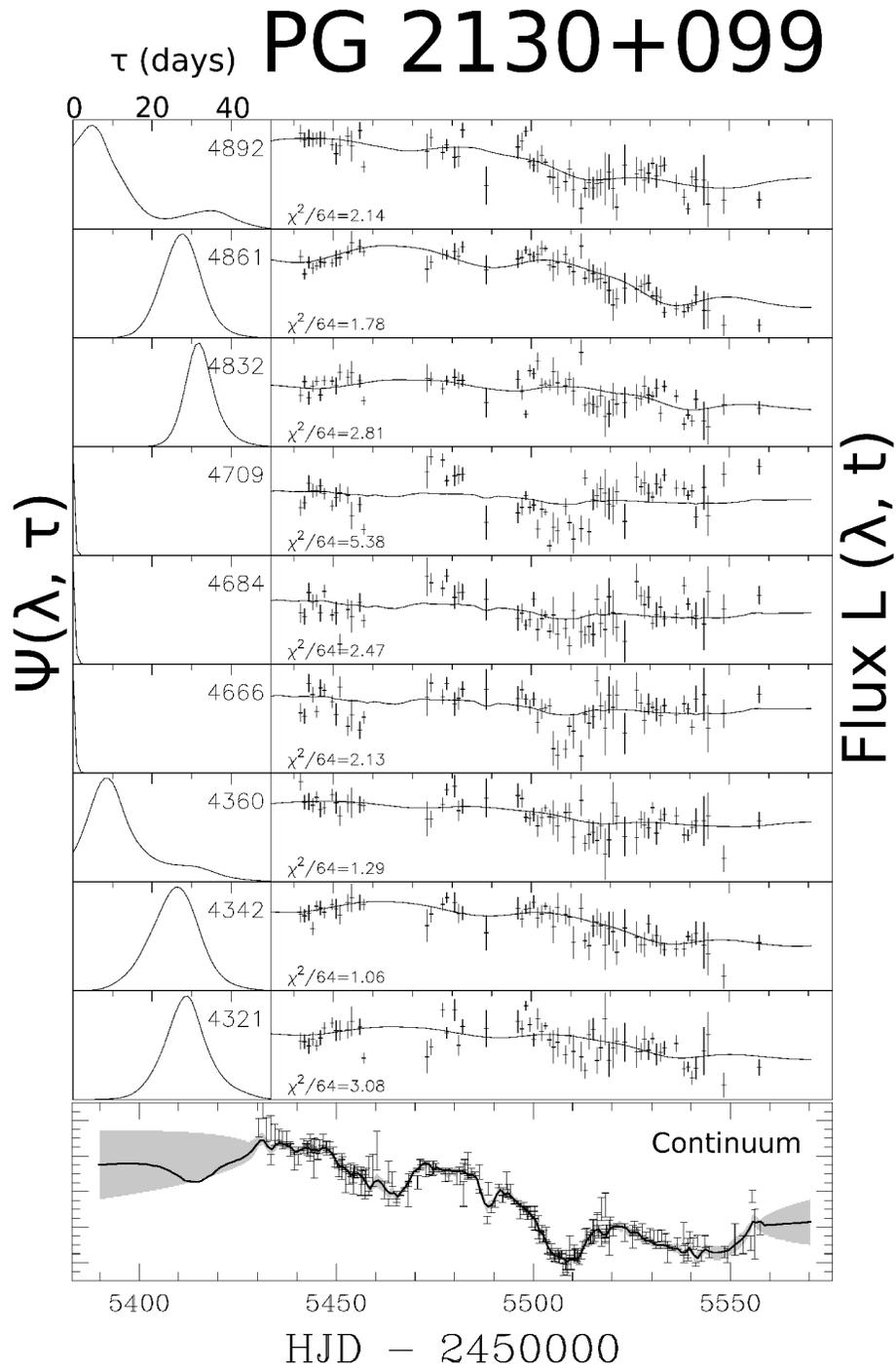}
\caption{Best MEMECHO fits to the spectra of PG\,2130+099. Panels and
symbols are the same as in Figure \ref{fig:mrk335fits}.}
\label{fig:pg2130fits}
\end{center}
\end{figure}

Using the JAVELIN mean continuum light curve allows us to fit the
variability of both the continuum and the rest of the spectrum to much
higher degrees of accuracy and keep MEMECHO from introducing spurious
features into our light curves and by extension, into the derived echo
maps. Light curves, 1-dimensional MEMECHO fits, and recovered delay
maps at selected wavelengths are shown in Figures
\ref{fig:mrk335fits}-\ref{fig:pg2130fits} for all five objects. The
wavelengths for the fits and delay maps shown in these figures were
chosen to show the response in the red and blue wings as well as at
the center of each emission line. One thing to note is that using the
probability-weighted mean as the constraint on the MEMECHO
construction of the continuum means that we are over-smoothing the
light curves because we are neglecting the covariance of light curve
deviations from the mean, which may weaken short-timescale
variability. This will also cause some difficulty in the fitting, as
it can force the light curve models to be smoother than the data. At
the same time, the $\chi^2/N$ value reported by MEMECHO is no longer
strictly valid --- the deviations of any particular light curve model
from the ``mean'' continuum light curve are highly correlated, so, for
example, having 10 consecutive points 1$\sigma$ from the light curve
model is likely a 1$\sigma$ deviation, not a 10$\sigma$
deviation. However, the velocity-delay maps do not change
significantly if we use random individual DRW realizations of the
light curve instead of the mean, indicating that the mean light curve
can be used to produce accurate velocity-delay maps. While this
solution is not ideal, it is currently the best method we have of
dealing with gaps in the observed continuum light curves pending a
major effort to integrate the two distinct software packages (JAVELIN
and MEMECHO).

\section{TWO-DIMENSIONAL VELOCITY-DELAY MAPS}
\label{sec:delaymaps}
We were able to recover velocity-delay maps with MEMECHO for four out
of the five galaxies observed. The best maps are for 3C\,120 and Mrk
335, while the ones obtained for PG\,2130+099 and Mrk 1501 are
somewhat less well-defined. The light curves for Mrk 6, on the other
hand, do not seem to be well-fit by a simple echo model, and we were
therefore unable to obtain two-dimensional velocity-delay maps for
this object. Figure \ref{fig:mrk6fits} shows the best model fits we
were able to obtain for Mrk 6. While there is some evidence for
velocity-dependent structure in the one-dimensional delay maps shown,
the models do not successfully fit any of the short-term variations
seen in this object. We were unable to improve the light curve fits by
lowering the target $\chi^2/N$, as MEMECHO was unable to converge on a
solution when we did so.

\begin{figure}
\begin{center}
\epsscale{0.4}
\plotone{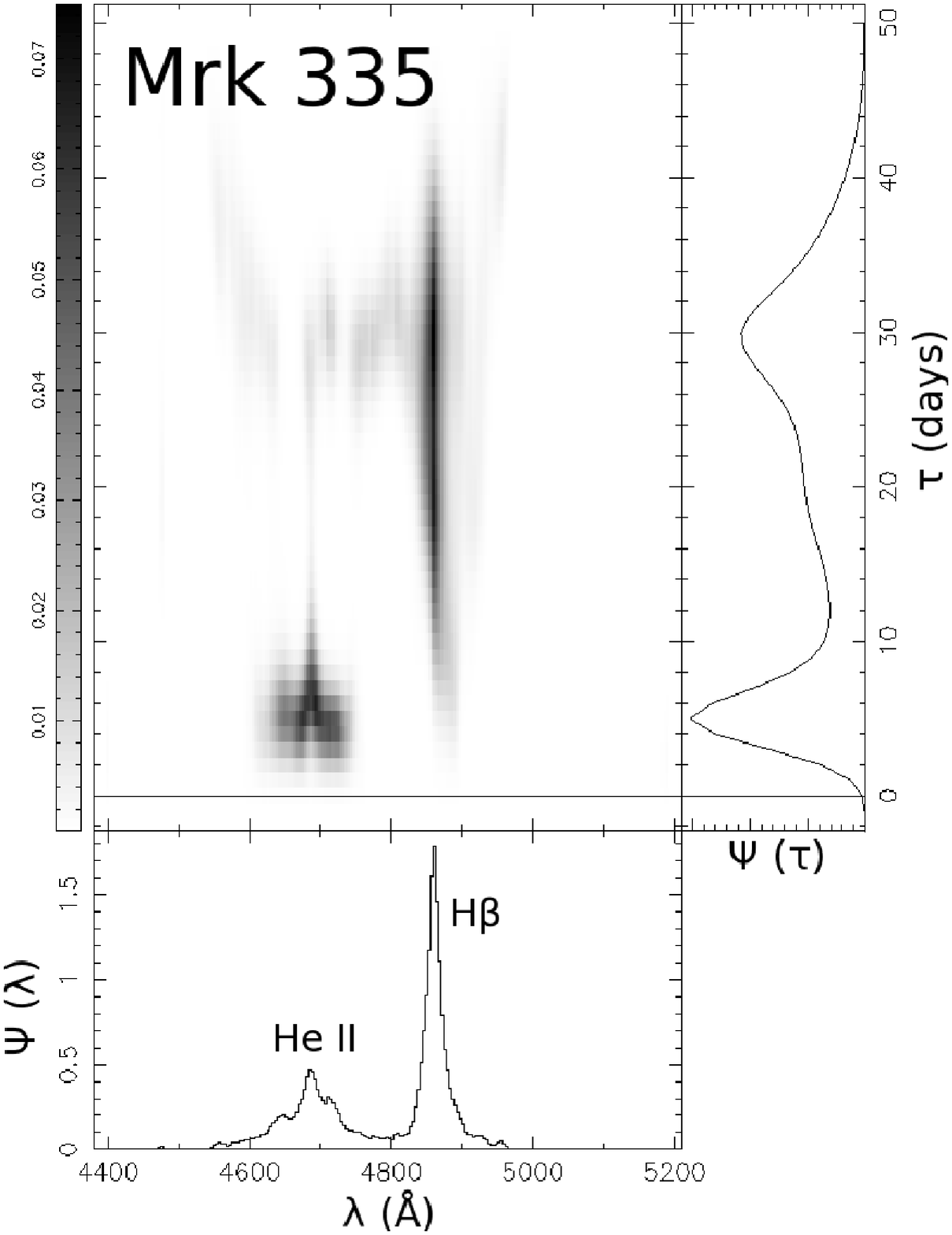}
\plotone{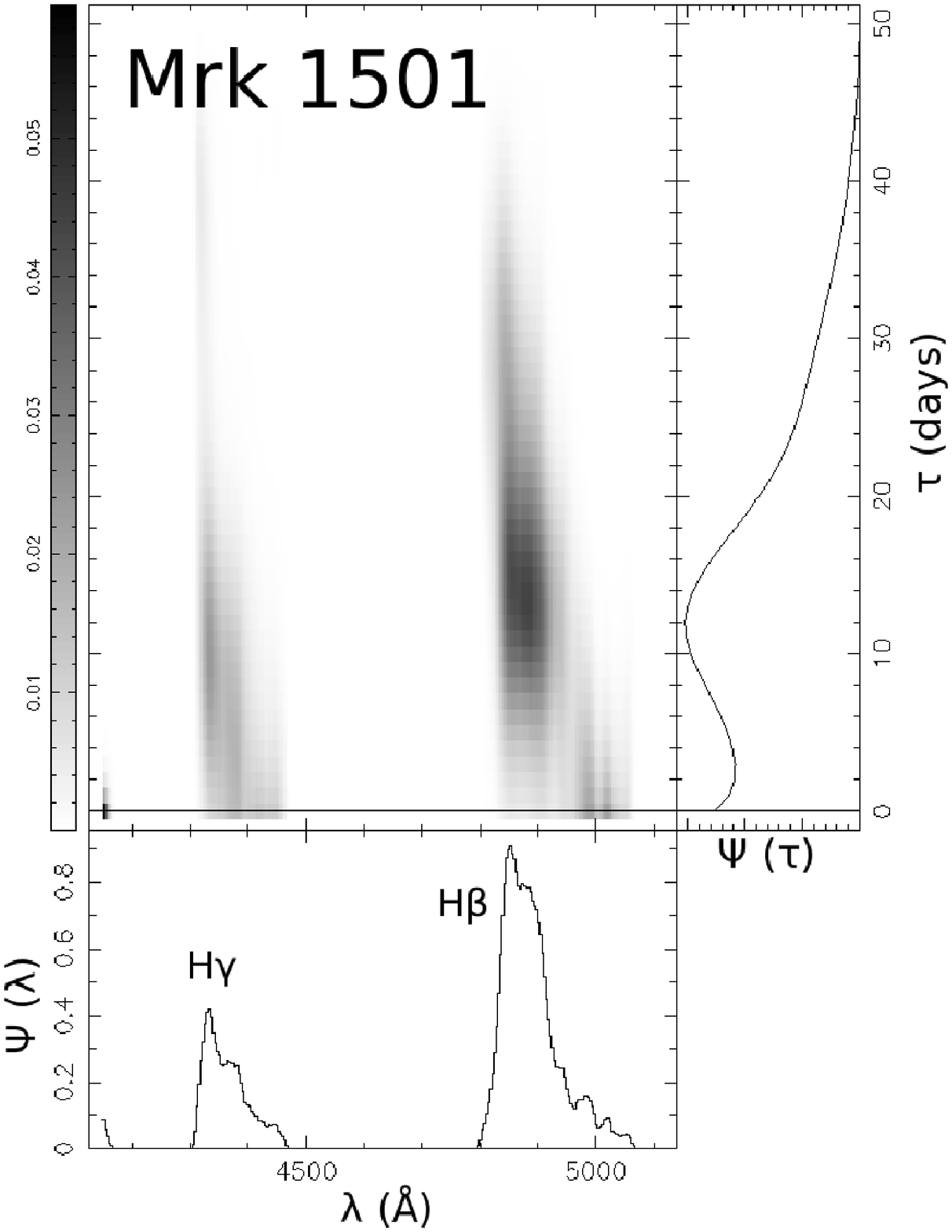}
\plotone{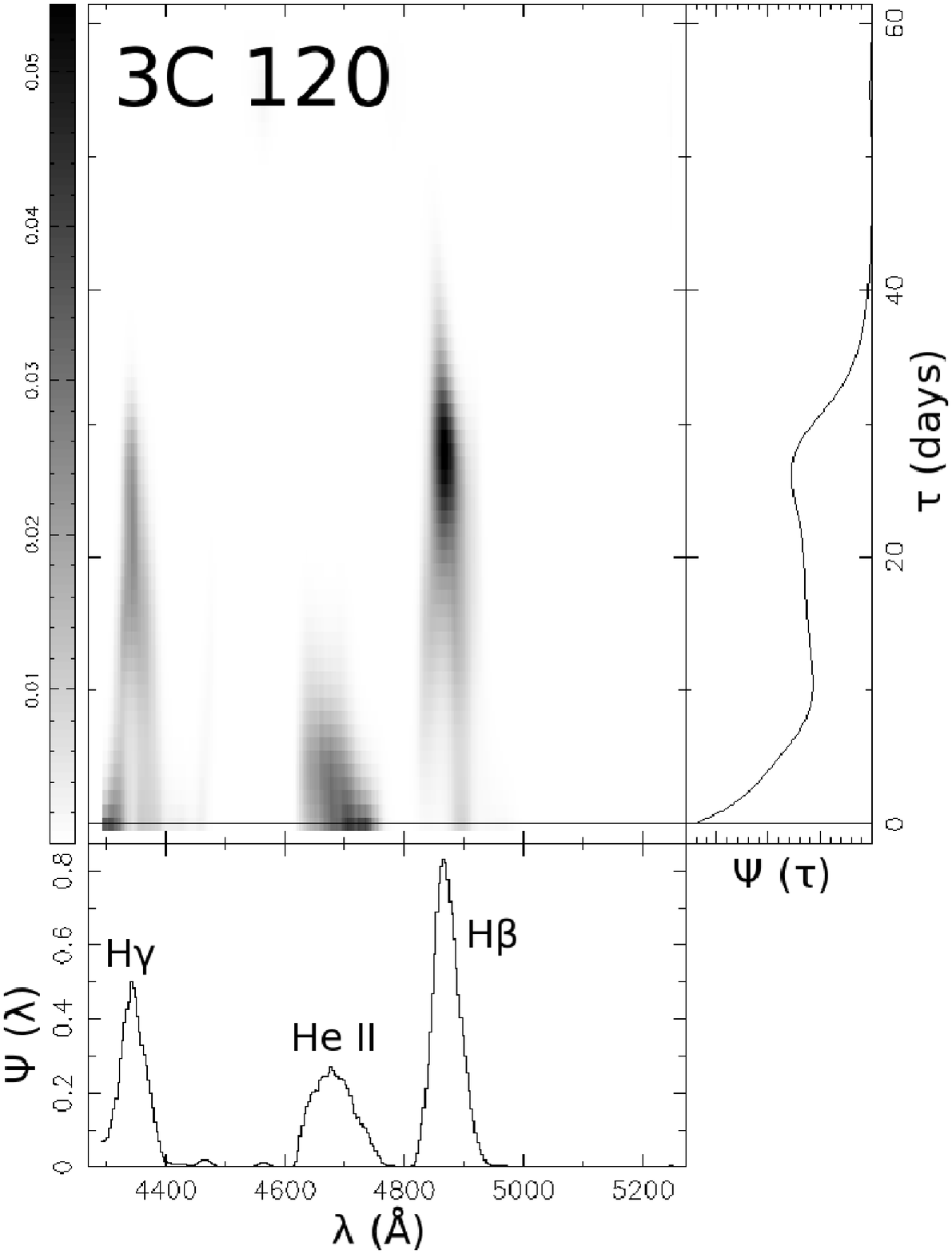}
\plotone{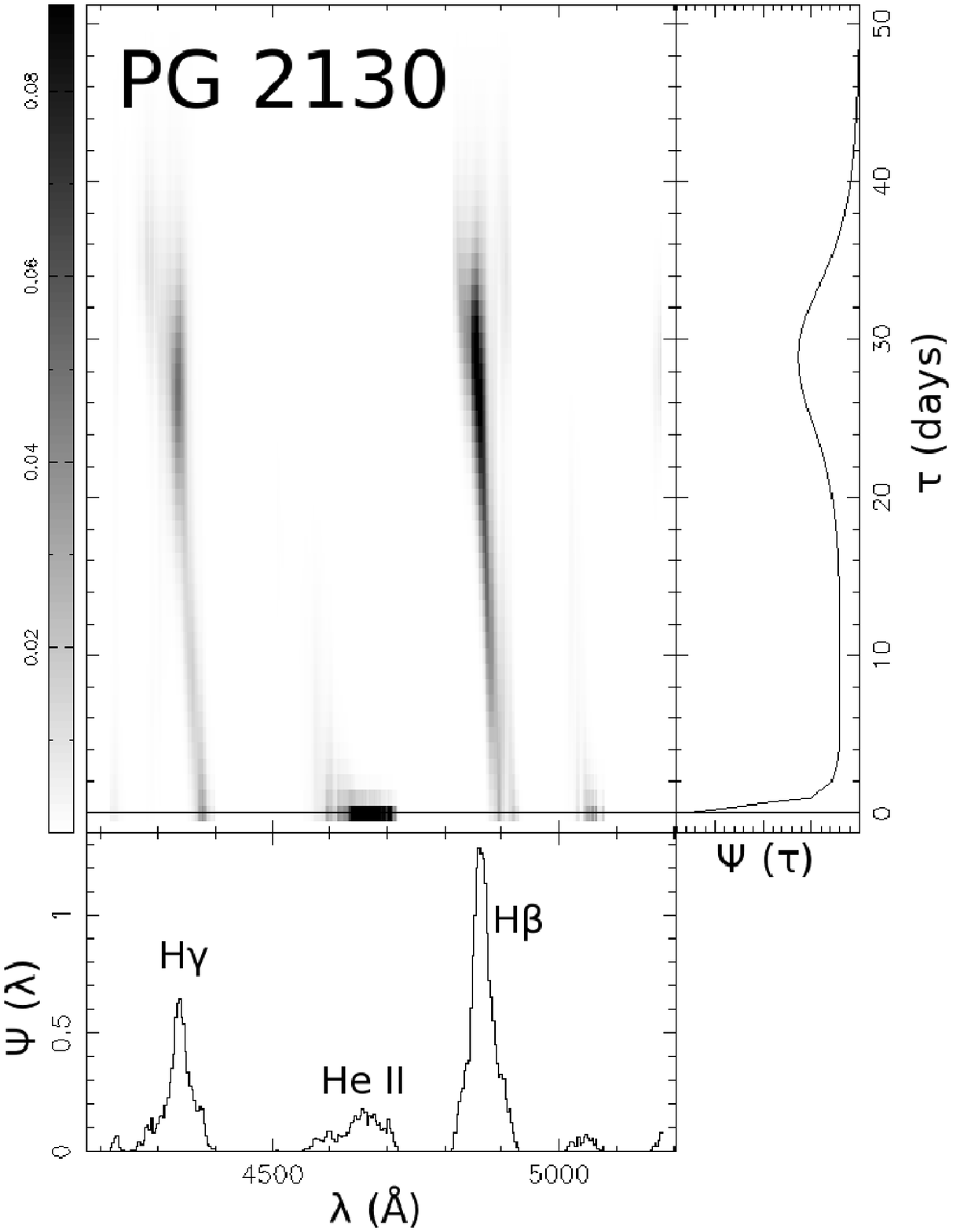}
\caption{Our best-fit velocity-delay maps over the full rest-frame
wavelength range for each object (grayscale), projections onto the
rest-frame wavelength axis (bottom panel) and the time-delay axis
(right panel). $\Psi(\lambda)$ is the overall response added up at
each wavelength, and $\Psi(\tau)$ is the overall response of all
emission lines added together at a given $\tau$.}
\label{fig:veldelay}
\end{center}
\end{figure}

\begin{figure}
\begin{center}
\epsscale{0.5}
\plotone{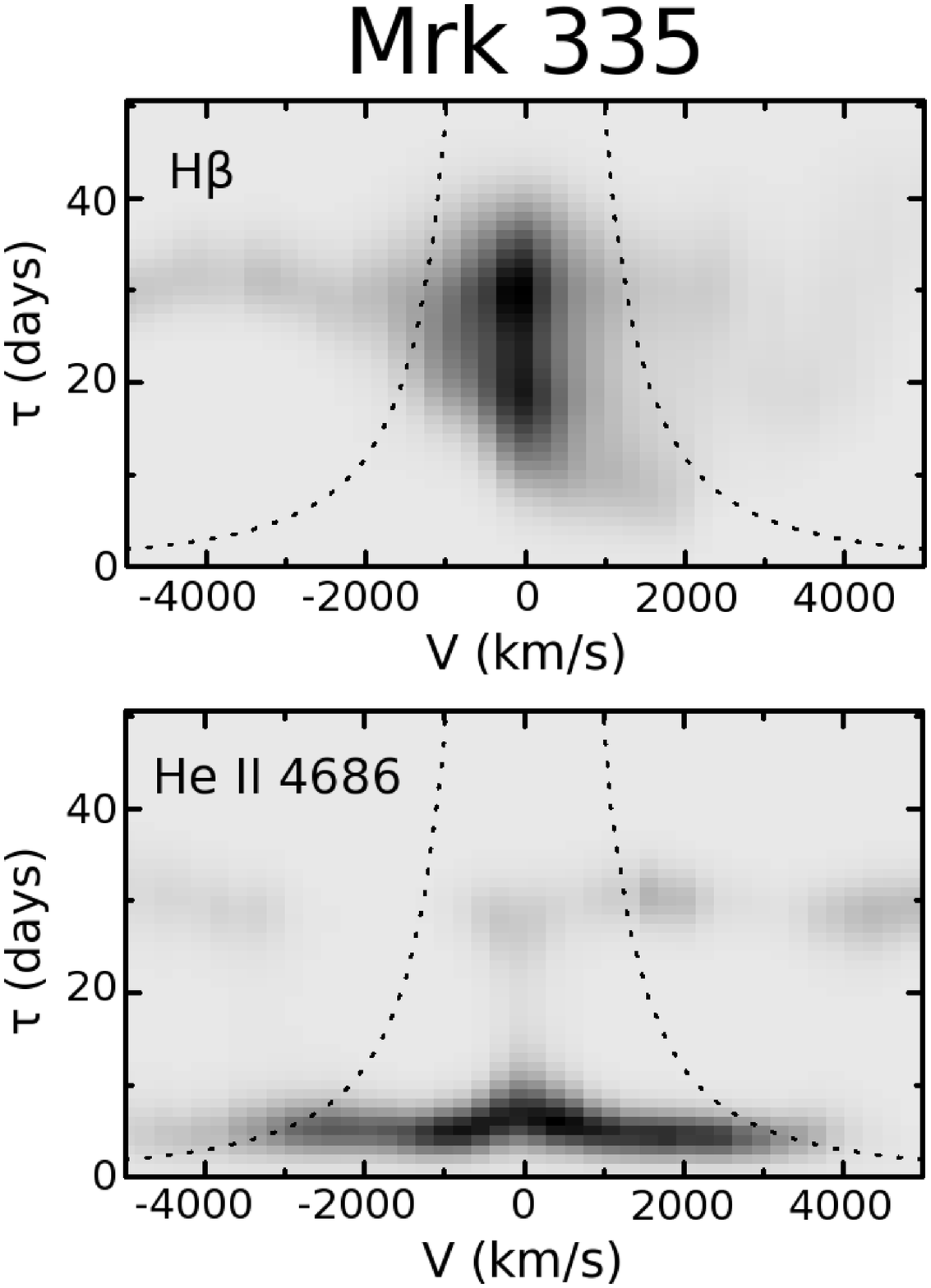}
\caption{Velocity-delay maps for both emission lines seen in Mrk
335. The dotted lines show the ``virial envelope'', $V^2\tau c/G = 4.6
\times 10^6$ \Msun, measured from the mean time lag (\citealt{Grier12b}).}
\label{fig:mrk335fancy}
\end{center}
\end{figure}

\begin{figure}
\begin{center}
\epsscale{0.5}
\plotone{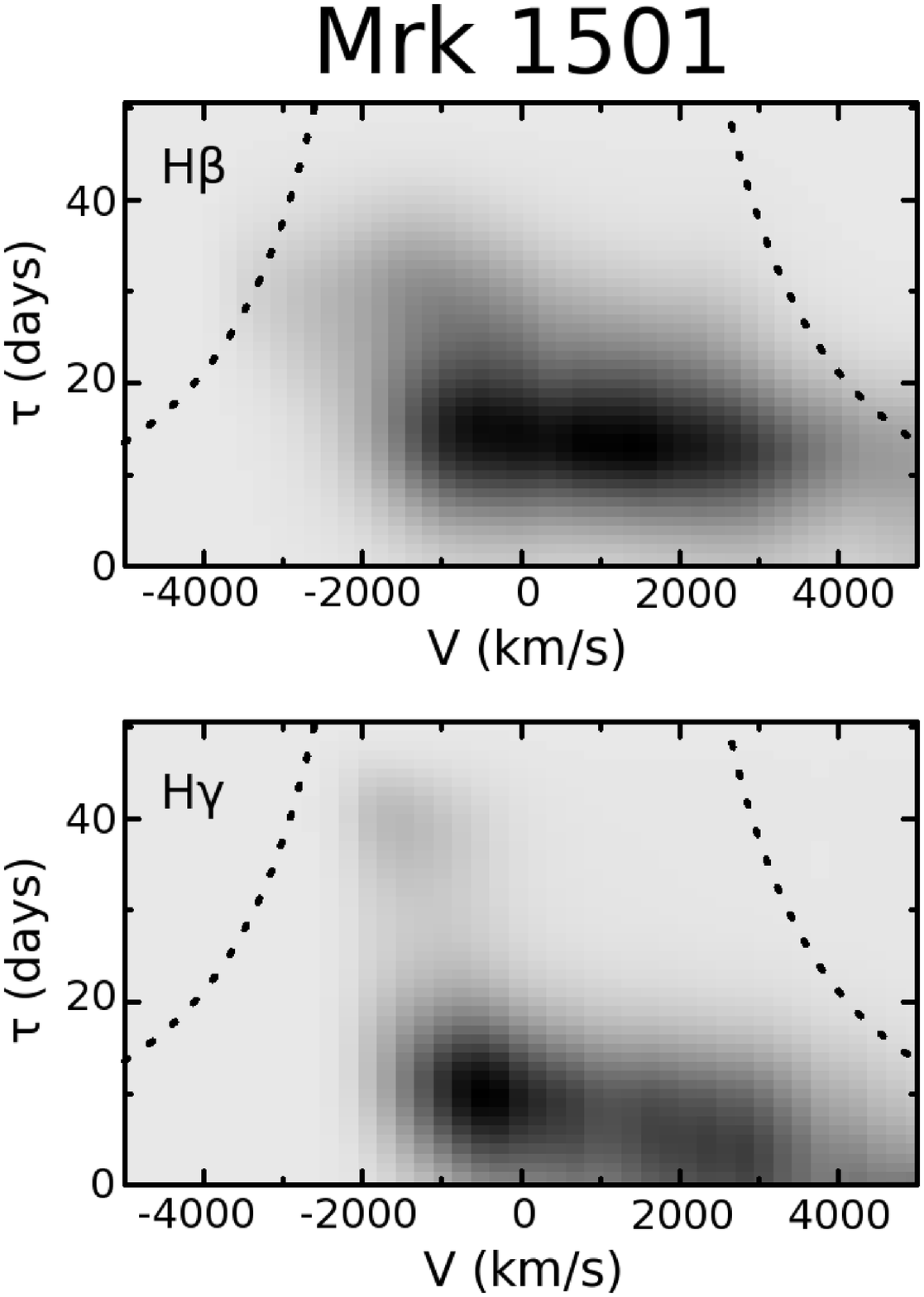}
\caption{Velocity-delay maps for both emission lines seen in Mrk
1501. The dotted lines show the ``virial envelope'', $V^2 \tau c/G =
3.3 \times 10^7$ \Msun, measured from the mean time lag (\citealt{Grier12b}).}
\label{fig:mrk1501fancy}
\end{center}
\end{figure}

\begin{figure}
\begin{center}
\epsscale{0.5}
\plotone{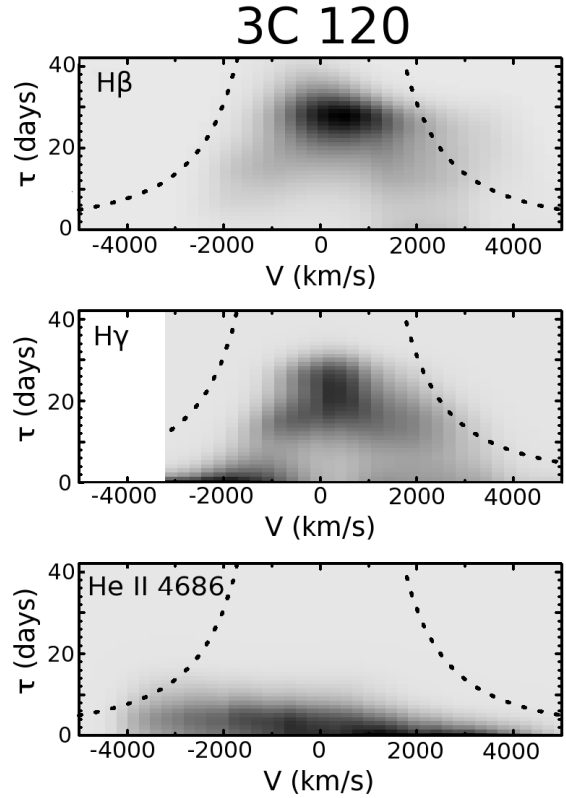}
\caption{Velocity-delay maps for all three emission lines seen in
3C\,120. The dotted lines show the ``virial envelope'', $V^2\tau c/G =
1.2 \times 10^7$ \Msun, measured from the mean time lag
(\citealt{Grier12b}). The velocity-delay map for \Hgamma \ is
truncated at V$\sim$3250 km/s because the wavelength range of our
spectrograph did not cover this far into the red. }
\label{fig:3c120fancy}
\end{center}
\end{figure}

The best-fit velocity-delay maps for the other four objects, covering
the full observed wavelength range for each, are shown in Figure
\ref{fig:veldelay}. The parameters used in MEMECHO to create the
velocity-delay maps are given in Table \ref{Table:MEMECHO}. We also
show more detailed velocity-delay maps for each individual emission
line recovered for each object in Figures
\ref{fig:mrk335fancy}-\ref{fig:pg2130fancy}. To aid the eye in
comparing ionization-stratified BLR structure between different
emission lines in each object, we also provide false-color maps in
Figure \ref{fig:colorplots}, with each color representing a different
emission line as described in the captions. We also created simulated
velocity-delay maps for a few different BLR kinematic models to
compare qualitatively with our recovered velocity-delay maps. These
simulated maps are shown in Figure \ref{fig:toys}, and represent
different BLR models around a black hole with \mbh = $1\times10^7$
\Msun. We see features reminiscent of these simple models in our
recovered velocity-delay maps.

\subsection{Comments on Individual Objects}
\subsubsection{Mrk 335}
\begin{figure}
\begin{center}
\epsscale{0.49}
\plotone{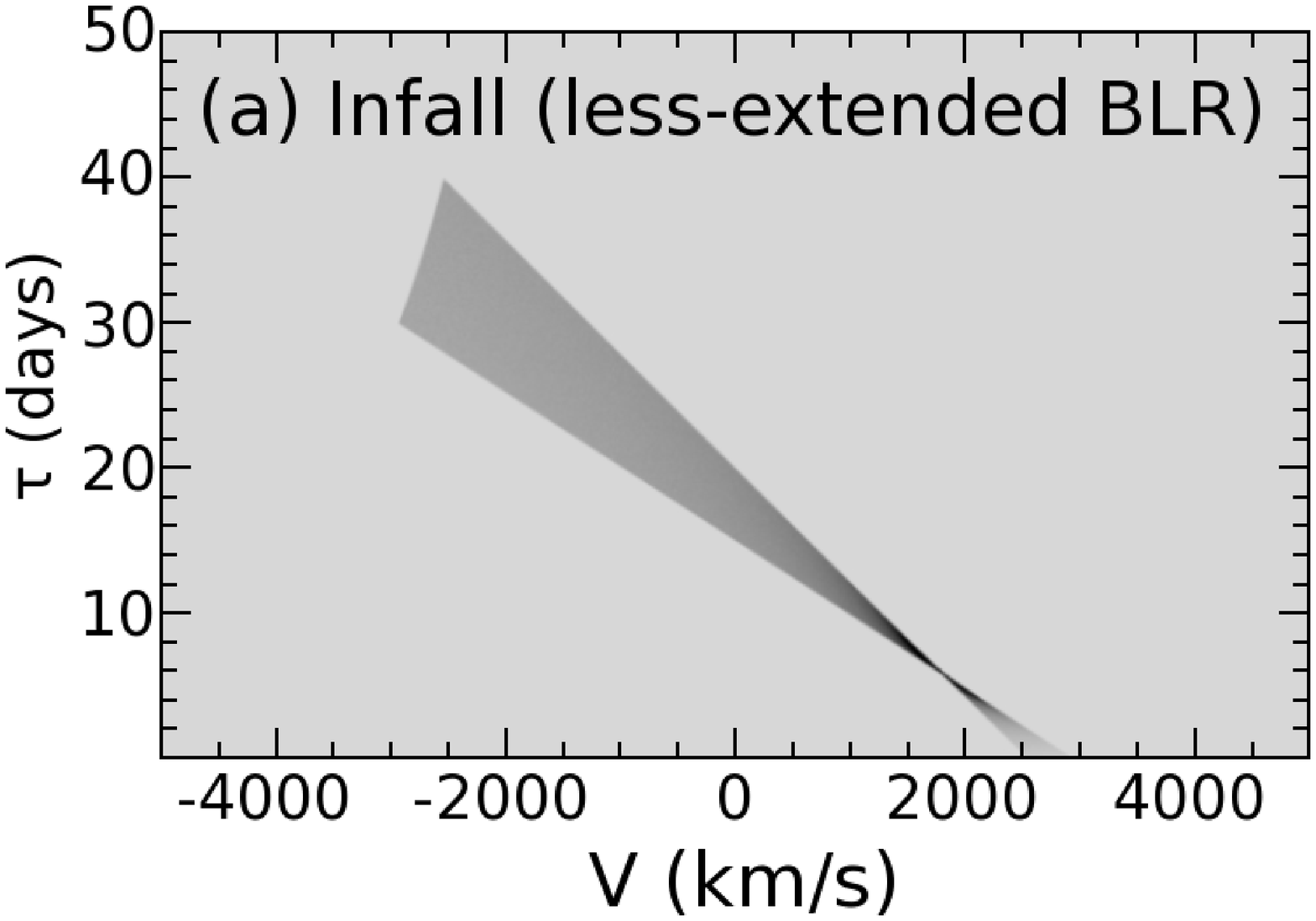}
\plotone{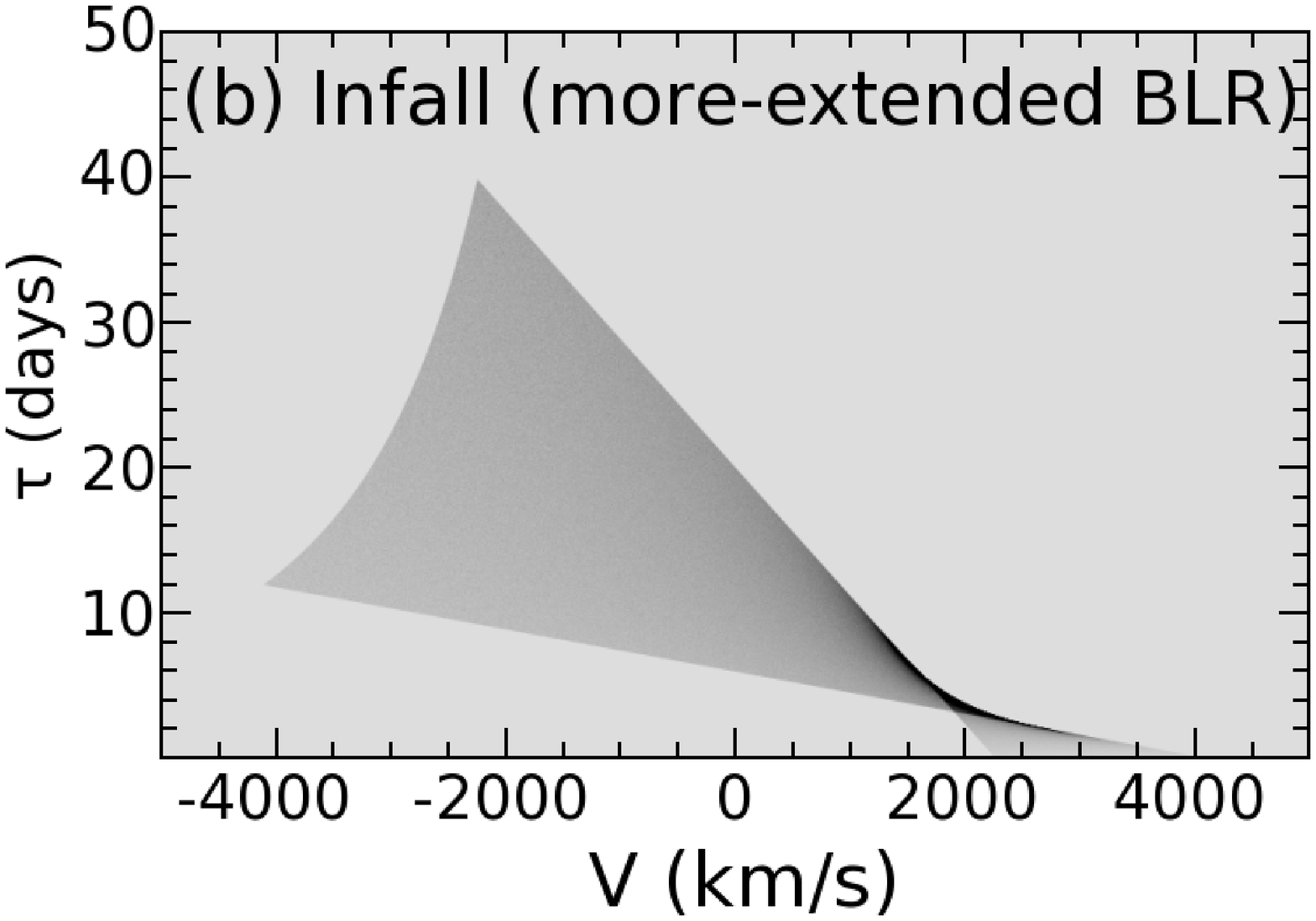}
\plotone{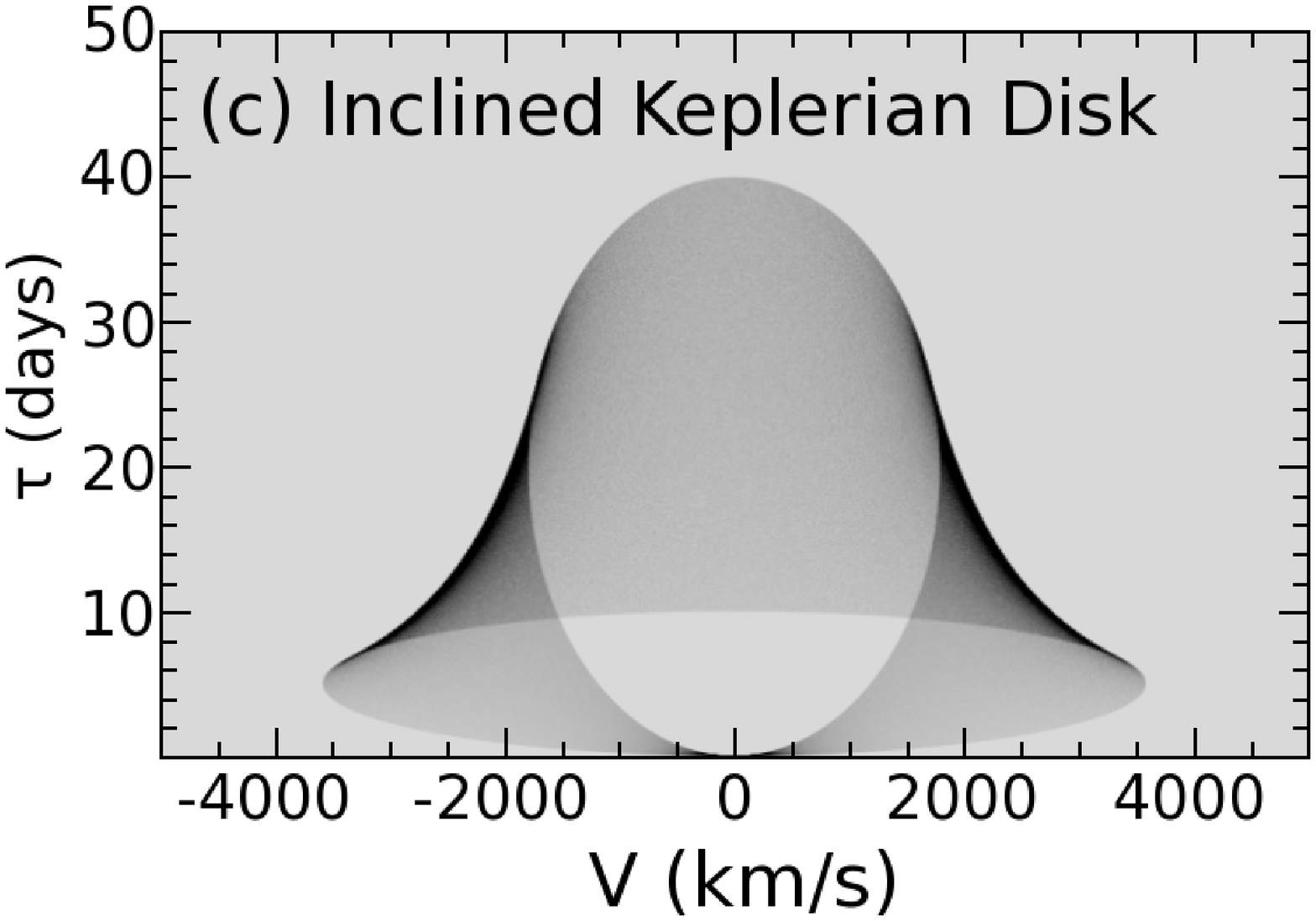}
\plotone{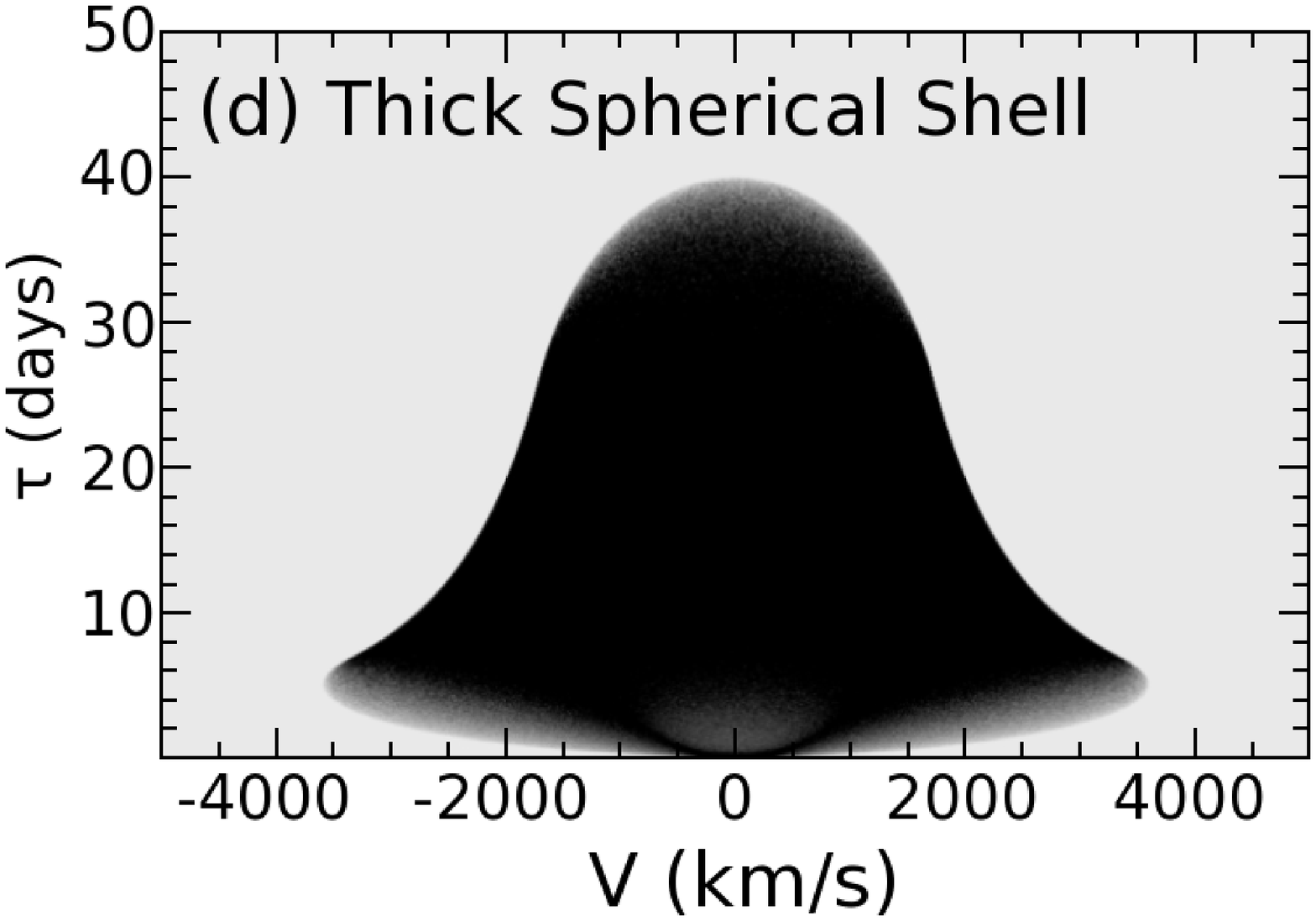}
\caption{Velocity-delay maps for simple BLR models of \Hbeta \
emission around a 1$\times$10$^7$ \Msun \ black hole. Panel (a) shows
highly beamed emission (typical for \Hbeta, see \citealt{Ferland92})
from gas in free-fall motion. The infalling gas is distributed in a
spherical shell with inner and outer radii of 15 and 20 light-days,
inclined at an angle of 45 degrees. Panel (b) shows the same infall
model but with a more-extended BLR with inner and outer radii of 5 and
20 light-days, respectively. Panel (c) shows a map for an edge-on
Keplerian disk with inner and outer radii of 5 and 20 light-days, and
panel (d) shows a map for a fully illuminated thick spherical shell of
Keplerian circular orbits, with inner and outer BLR radii of 5 and 20
light-days.}
\label{fig:toys}
\end{center}
\end{figure}

The velocity-binned analysis of Mrk 335 shows a definite
velocity-dependent lag signature (Figure \ref{fig:velres}), and we
clearly see similar structures in the velocity-delay maps of both the
\heii \ and \Hbeta \ emission lines (Figure~\ref{fig:mrk335fancy}). We
see a chevron-shaped pattern in the \heii \ line, with a lack of
prompt response in the center and shorter delays in the wings, which
is consistent with the signatures expected from an inclined disk or a
spherical shell (Figure \ref{fig:toys}). In the \Hbeta \ emission, we
see an asymmetric profile, with longer lags towards the blue end of
the emission, and shorter lags towards the red end. This is suggestive
of inflowing gas, as demonstrated in our simple infall models given in
Figure \ref{fig:toys}. This also matches the signature we found in our
initial velocity-binned analysis. The \heii \ emission is confined to
smaller delays, as is expected from photoionization models of the
BLR. Also consistent with disk structure is the response of \heii \
along a much wider velocity range than that of \Hbeta.

\subsubsection{Mrk 1501}
Fewer kinematic details are apparent in the recovered maps for Mrk
1501 than for the other objects in our sample. This may be a
consequence of the noisier continuum and emission-line light curves
obtained for this target. Nonetheless, we can still gain useful
insights into the \Hbeta \ and \Hgamma-emitting regions of the BLR in
this object. As with Mrk 335, we see evidence in the velocity-delay
maps for inflow, with longer lags in the blue and shorter lags towards
the red end of the \Hbeta \ and \Hgamma \ emission lines. This
signature is closest to that of the ``extended BLR'' infall model
shown in Figure \ref{fig:toys}. While this velocity-delay map is
probably too low-resolution for any detailed modeling, we do see this
same signature in our velocity-resolved analysis in
Figure~\ref{fig:velres}. There is also evidence for radial
stratification of the BLR --- Figure~\ref{fig:colorplots} shows there
is a much stronger \Hbeta \ response at longer time lags than is
observed for the H$\gamma$ \ emission. This is consistent with the
idea that photoionization physics regulates the size of the BLR, as
\Hgamma \ is a higher-energy transition and would therefore need to be
located closer to the ionizing source than \Hbeta.

\subsubsection{3C\,120}
Our velocity-delay maps for 3C\,120, shown in full in
Figure~\ref{fig:veldelay}, are the cleanest maps we were able to
recover. We see evidence for radial stratification, with \heii \
showing the shortest delays and \Hgamma \ and \Hbeta \ emitted at
progressively larger radii. Figure~\ref{fig:colorplots} shows all
three emission lines on one scale, with \Hbeta, \Hgamma, and \heii \
shown in red, blue, and green, respectively, highlighting the radial
stratification. The lack of a prompt response at line center in both
Balmer lines indicates a deficit of material along the line of
sight. The shape of the hydrogen response is consistent with
signatures expected from an inclined disk or a spherical shell (Figure
\ref{fig:toys}). This is similar to the signal found in NGC5548
(\citealt{Horne91}), Arp 151 (\citealt{Bentz10b}), and Mrk 50
(\citealt{Pancoast12}). We also see an asymmetry in the strength of
the response in both the \heii \ and \Hbeta \ emission profiles,
nominally indicating inflow. The smaller velocity range of the \Hbeta
\ and \Hgamma \ emission with respect to \heii, combined with the
radial stratification signatures, is consistent with disk structure.

\subsubsection{PG\,2130+099}
The velocity-delay map for PG\,2130+099 is somewhat noisier than that
of 3C\,120, and the model fits are not as good. However, the structure
of the map is worth noting. While reverberation delays for the \heii \
emission remain unresolved for this data, both the \Hbeta \ and
\Hgamma \ emission lines clearly show velocity-resolved delay
structure, with strong asymmetries as a function of velocity across
the emission-line profile. This is similar to the \Hbeta \ emission we
see in Mrk 335 and the \heii \ emission in 3C\,120. This asymmetry,
with longer lags at the blue end of the emission line and shorter lags
to the red, is suggestive of inflowing gas, and matches the ``Infall
(less-extended BLR)'' model in Figure \ref{fig:toys} quite well. In
our analysis of the velocity-binned sections of the \Hbeta \ emission
line (Figure \ref{fig:velres}), we see this same structure, so while
the delay maps for this object are probably not good enough for any
detailed modeling, they are consistent with the signatures of
inflowing material already seen in this object.

\begin{figure}
\begin{center}
\epsscale{0.5}
\plotone{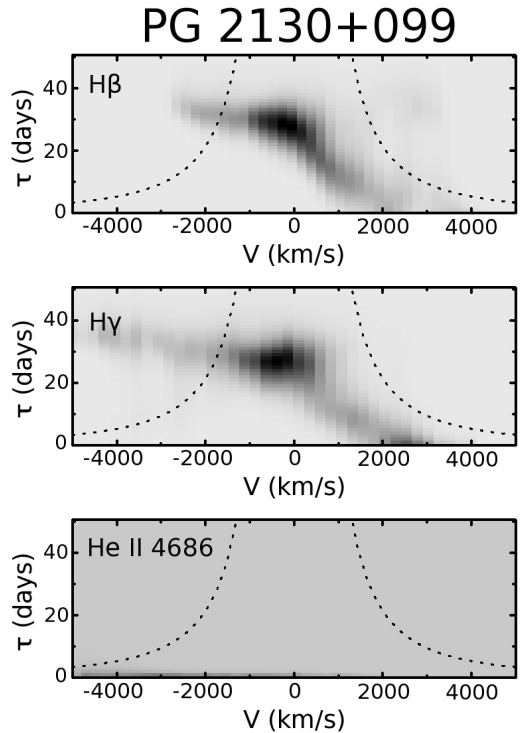}
\caption{Velocity-delay maps for all three emission lines seen in
PG\,2130+099. The dotted lines show the ``virial envelope'', $V^2\tau
c/G = 8.3 \times 10^6$ \Msun, measured from the mean time lag
(\citealt{Grier12b}).}
\label{fig:pg2130fancy}
\end{center}
\end{figure}

\begin{figure}
\begin{center}
\epsscale{0.49}
\plotone{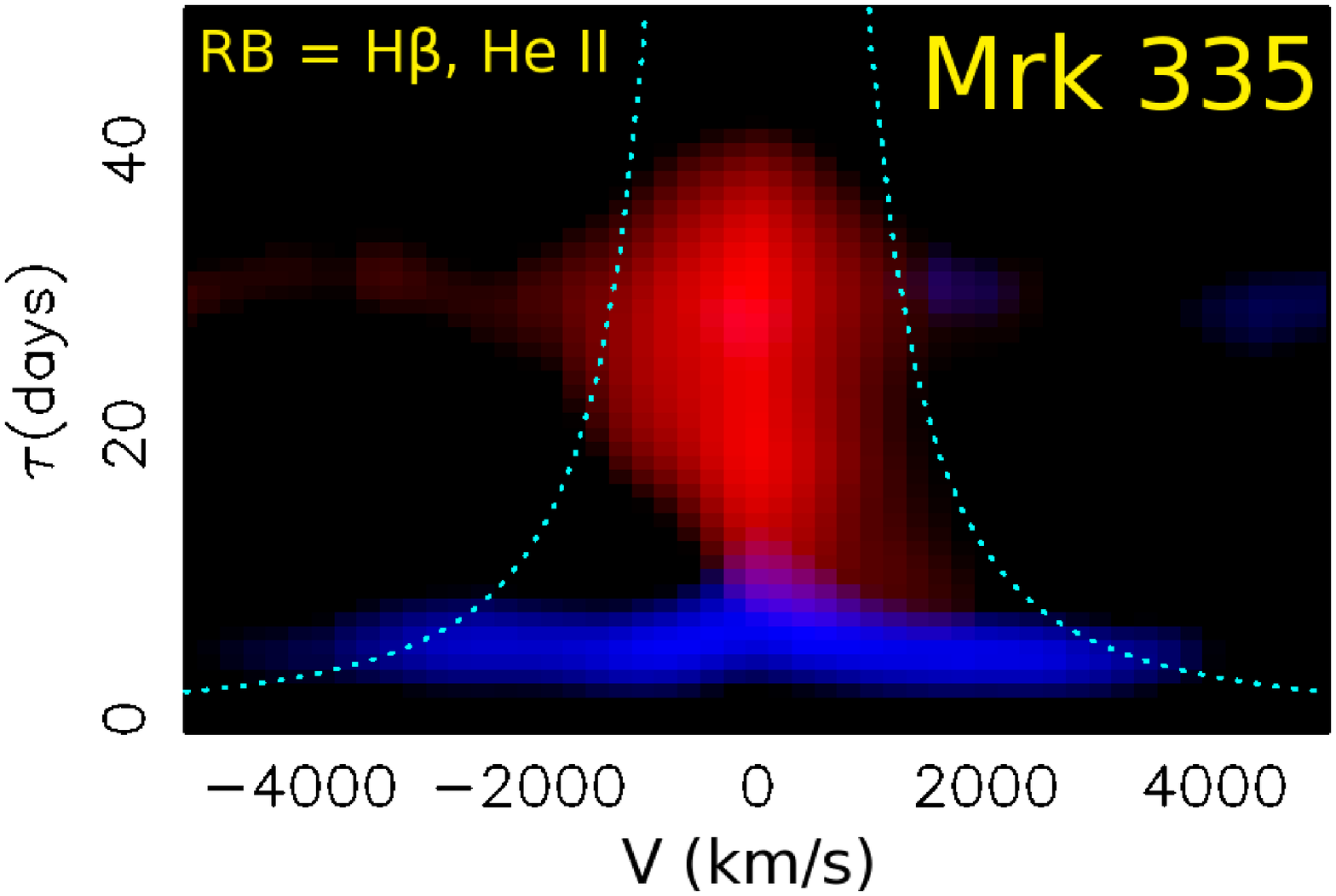}
\plotone{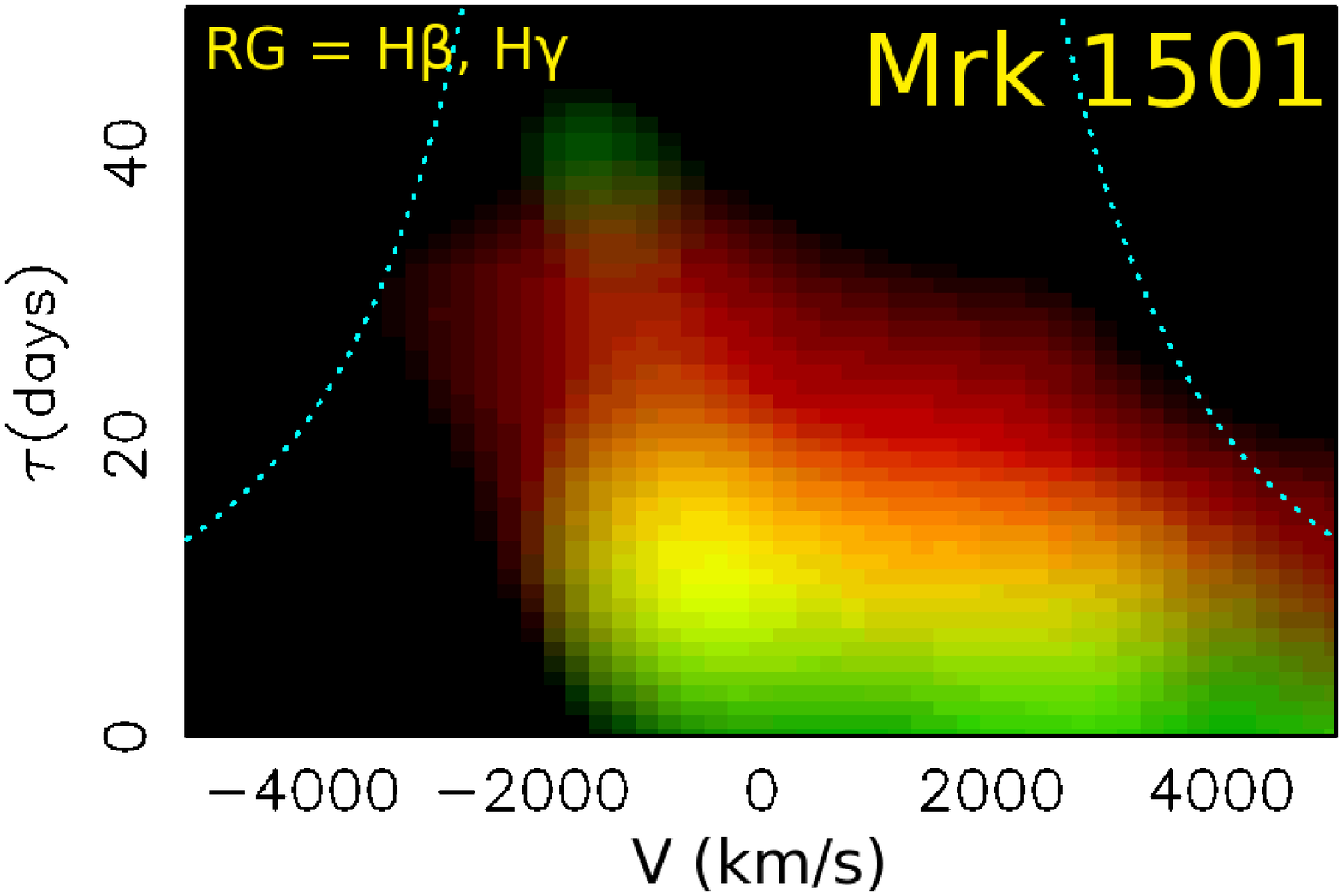}
\plotone{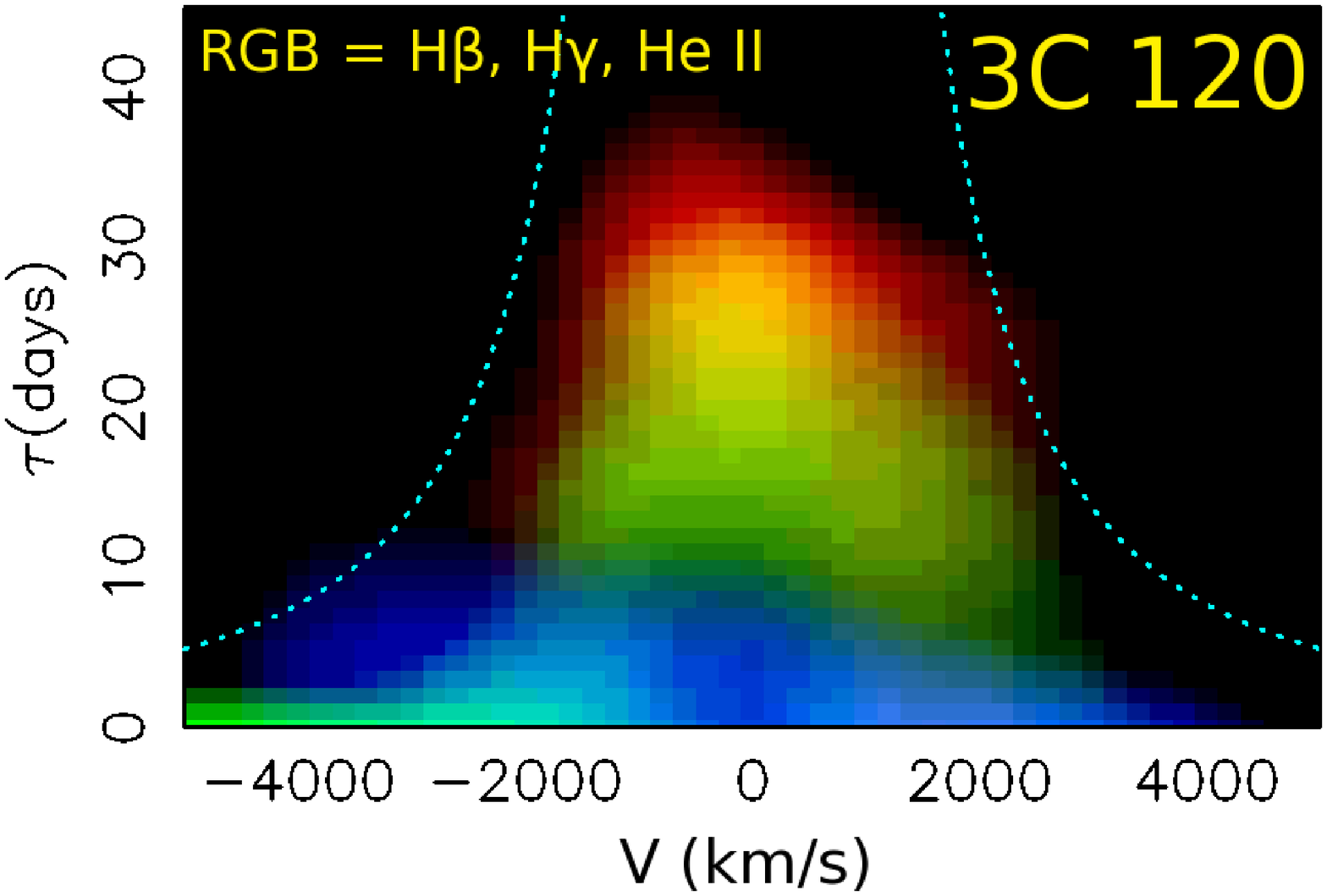}
\plotone{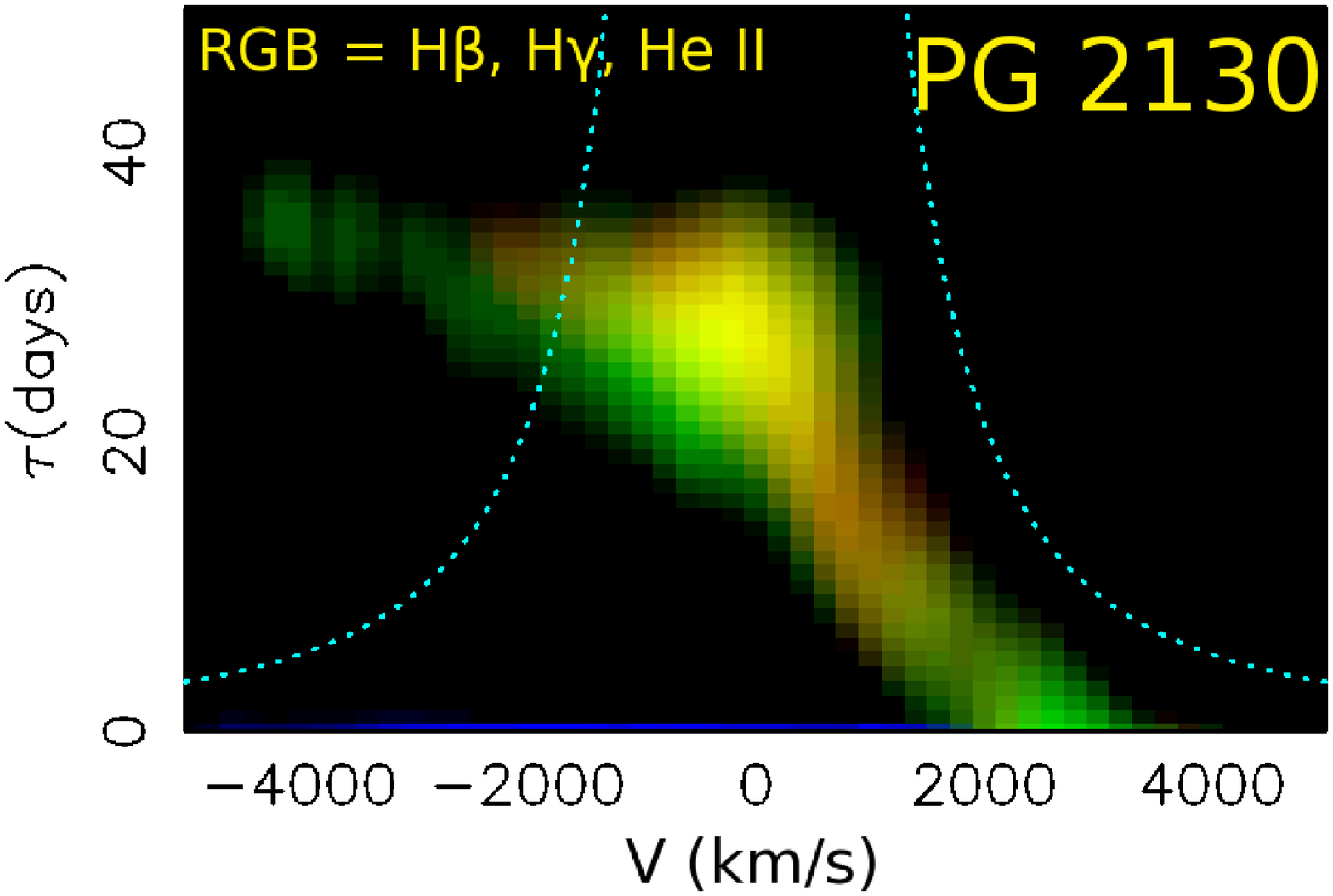}
\caption{False-color velocity-delay maps for Mrk 335, Mrk 1501,
3C\,120, and PG\,2130+099. The dotted lines in each panel correspond
to virial envelopes for each object as listed in Figures
\ref{fig:mrk335fancy}-\ref{fig:pg2130fancy}. The \Hbeta \ emission is
shown in red, \Hgamma \ emission in green, and \HeII \ emission in
blue.}
\label{fig:colorplots}
\end{center}
\end{figure}

PG\,2130+099 has long been a curiosity. Early on, \cite{Kaspi00}
measured a time lag of $\sim$180 days in this object, placing it well
above the \radlum \ relationship. Later studies by \cite{Grier08} and
\cite{Grier12b} found much shorter lags on the order of tens of days:
Most recently, \cite{Grier12b} reported a mean \Hbeta \ time delay of
12.8$^{+1.2}_{-0.9}$ days. These studies attributed the discrepancies
to undersampled light curves combined with long-term secular changes
in the \Hbeta \ equivalent width in the data from
\cite{Kaspi00}. However, even with the new, shorter lag measurements,
PG\,2130+099 is still a major outlier from the \radlum \ relation, as
it is now positioned far below the relation
(\citealt{Grier12b}). Despite the higher sampling rate of the more
recent campaigns, ambiguities remain as to whether the measured \Hbeta
\ lags represent the true mean BLR radius, as the light curves were
missing data at key points in time. We see in our velocity-delay map
that the majority of the response in the \Hbeta \ emission seems to be
centered on a delay of $\sim$30 days (Figures \ref{fig:pg2130fancy}
and \ref{fig:colorplots}).

\begin{figure}
\begin{center}
\epsscale{0.8}
\plotone{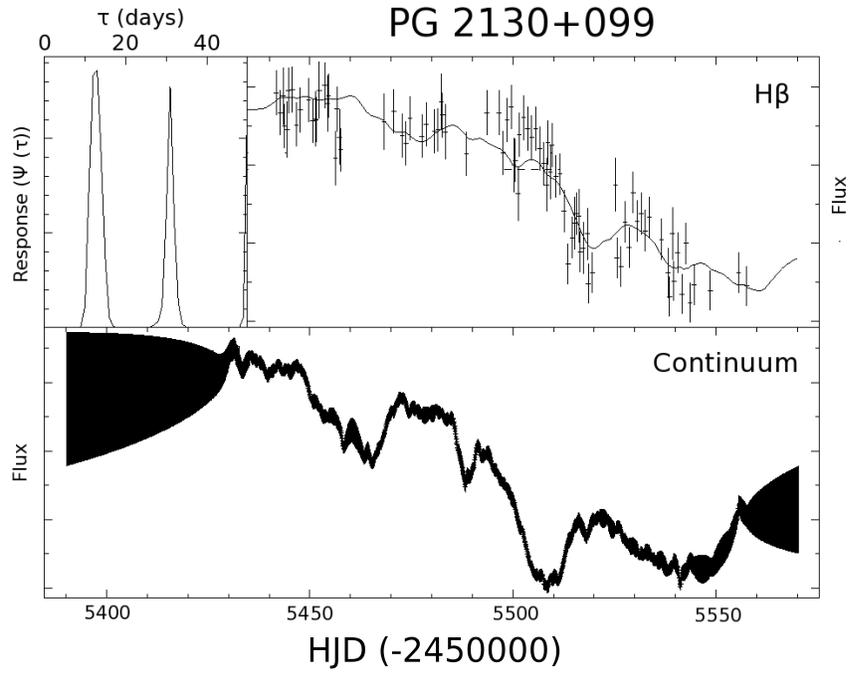}
\caption{One-dimensional delay map for PG\,2130+099. The bottom panel
shows the simulated continuum light curve used in the MEMECHO
analysis, with the errors shown as the black envelope. The top right
panel shows the light curve for the entire \Hbeta \ emission line from
\cite{Grier12b}. The top left panel shows the one-dimensional delay
map from MEMECHO.}
\label{fig:onedpg2130}
\end{center}
\end{figure}

To investigate this, we ran a one-dimensional delay map analysis of
PG2130\,+099 in MEMECHO to look for an indication of where the true
lag lies. Figure \ref{fig:onedpg2130} shows the model continuum light
curve envelope in the bottom panel, and the \Hbeta \ light curve from
\cite{Grier12b} in the top right panel; the top left panel shows the
delay map recovered by MEMECHO. The MEMECHO model fits the data fairly
well in this case, and there are two clear peaks in the delay map. The
stronger peak is centered around 12.5 days, and the slightly weaker
peak is centered at 31 days. We compare this with the two-dimensional
velocity-delay map (Figures \ref{fig:veldelay}, \ref{fig:pg2130fancy},
and \ref{fig:colorplots}), which shows a large signal on the blue side
of the emission concentrated at around 30 days and a fainter signal to
the redward side that stretches down to shorter lags. A plausible
model reproducing these results is a nearly face-on disk with the
emitting gas located at around 30 light-days, combined with a strong
inflowing gas component not necessarily within the plane of the
disk. Including an inflow signature when we measure the flux of the
entire \Hbeta \ emission line could cause our result to be skewed
towards shorter mean lags, when the true distance of the virialized
gas is closer to $\sim$30 days. Because of the lower quality and
coarser sampling of the light curves for this object, we will likely
be unable to model this structure in much more detail. However, it is
clear from the delay map that the majority of the \Hbeta \ signal
comes from a radius of $\sim$31 light-days. This radius puts
PG2130+099 much closer to the \radlum relation. This also increases
the black hole mass estimate for PG\,2130+099 by a factor of about
2.4, putting it at about 10$^8$ \Msun. This would place PG\,2130+099
definitely within the expected scatter of the \msigma relation.

\subsection{MEMECHO Parameters and Settings}
\subsubsection{The $A$ and $W$ Parameters}
\label{sec:aandw}
MEMECHO gives the user control over various aspects of the fitting
process. We first consider the weights $w_i$ (see Equation
\ref{eq:entropy}), which are implemented in
MEMECHO as the user-controlled $W$ and $A$ parameters. The $W$
parameter controls the weight given to pixels in $\Psi(\lambda,\tau)$
relative to the weight given to the pixels in the continuum model
$C(t)$. Increasing $W$ makes the delay map stiffer and allows for more
flexibility in the continuum model. $A$ affects the aspect ratio of
features in the velocity-delay map $\Psi(\lambda,\tau)$. The default
values for $\Psi(\lambda,\tau)$ are geometric means of neighboring
pixels, with $A$ increasing the weight of neighbor pixels in the
$\lambda$ direction relative to those in the $\tau$ direction.

\begin{figure}
\begin{center}
\epsscale{0.8}
\plotone{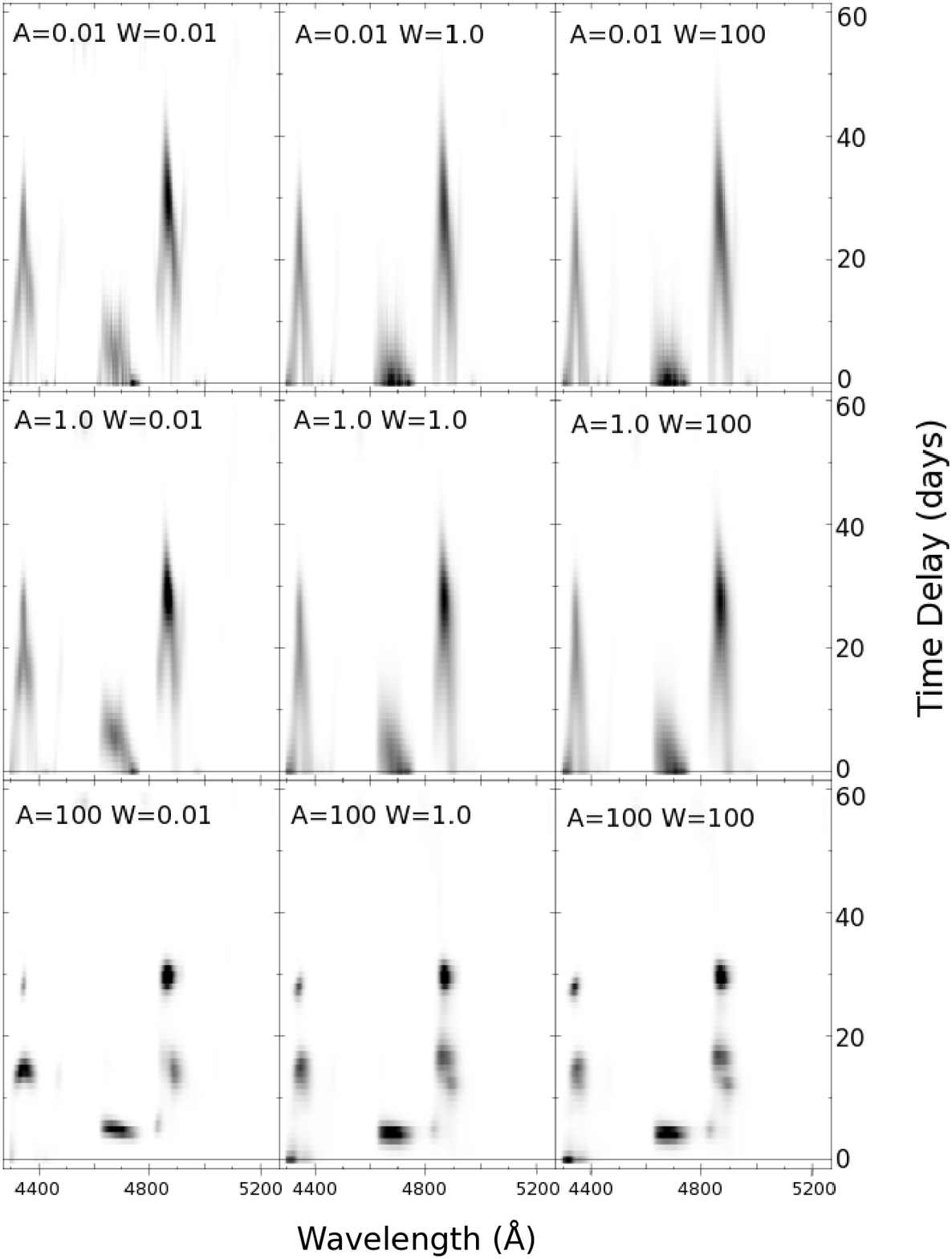}
\caption{Velocity-delay maps for 3C\,120. Each panel shows the best
map found as we vary $A$ and $W$ by factors of 100. Increasing $A$
smooths the maps more strongly in $\lambda$ and less in $\tau$, while
increasing $W$ reduces the importance of fitting the continuum model
relative to the delay maps.  All models are converged to the same
overall goodness-of-fit $\chi^2/N$. Our adopted velocity-delay map is
that of the middle panel.}
\label{fig:awmosaic}
\end{center}
\end{figure}

It is important that the structures we see in our delay maps are not
dependent on our choices of $A$ and $W$. To verify that our results
are robust, we produced MEMECHO models altering the $A$ and $W$
parameters and changing the wavelength and velocity grids of the
transfer function models. The panels of Figure \ref{fig:awmosaic} show
the resulting velocity-delay maps for 3C\,120 as we vary these
parameters. For the most part, the general shape of our velocity-delay
map is not affected by changes in $A$ or $W$. However, if $A$ is made
too large, we start to over-fit the data and introduce spurious
features into the maps. The stability with respect to the $W$
parameter is expected, since the continuum light curve we are using is
very highly sampled, and therefore very highly constrained. Since
these behaviors are generic to all four objects, we only show the
results for 3C\,120.

\subsubsection{The Minimum $\chi^2/N$}
We also investigated the effect of varying the target $\chi^2/N$ of
the MEMECHO solution, as the final resolution of the recovered delay
maps is controlled by this parameter. The choice of the target
$\chi^2/N$ is a trade off between smoother, lower delay resolution
maps for higher $\chi^2/N$, and sharper but less reliable structure in
the maps at lower $\chi^2/N$. Because we are using over-sampled
continuum model light curves from JAVELIN, the actual value of
$\chi^2/N$ is no longer strictly valid, because the continuum light
curve data points are not fully independent. Therefore, our chosen
values of $\chi^2/N$ reflect the best trade-off between delay-map
resolution and reliable structure in the fits that we could
obtain. Figure \ref{fig:ctmosaic} shows velocity-delay maps for
3C\,120 as we vary the $\chi^2/N$, from left to right. As we lower the
target $\chi^2/N$, the structure eventually becomes more complex but
also less reliable, as the fits are now producing structures to model
the noise in the line light curves. However, the basic structure of
the velocity-delay maps is robust to reasonable changes in the target
$\chi^2/N$.

\subsubsection{Minimum Allowed Time Delays}
\begin{figure}
\begin{center}
\epsscale{0.8} 
\plotone{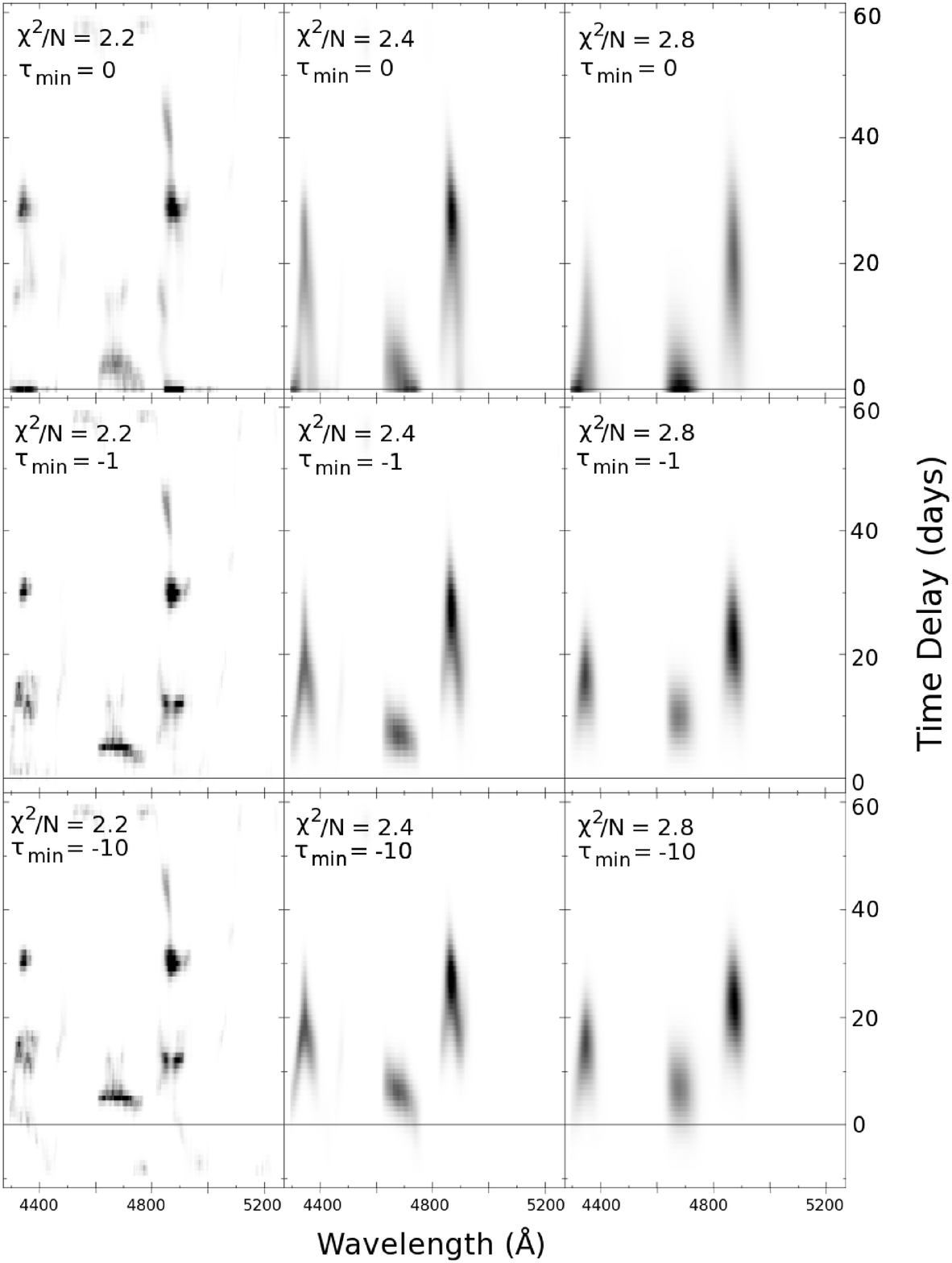}
\caption{Velocity-delay maps for 3C\,120, varying the degree of
smoothing and the minimum lag allowed in the model ($\tau_{\rm
min}$). Increasing values of $\chi^2/N$ mean that the entropy term is
more heavily smoothing the map. Our adopted velocity-delay map is that
of the top middle panel.}
\label{fig:ctmosaic}
\end{center}
\end{figure}

When we vary the minimum lag $\tau_{\rm min}$ that we allow MEMECHO to
consider, we also see a trend worth noting. Figure \ref{fig:ctmosaic}
shows that when $\tau_{\rm min}$ is set to zero (top panels), the
velocity-delay structure in the map also extends all the way to
zero. However, when negative lags are allowed (middle and bottom
panels), we see that the structure in the velocity-delay maps,
particularly when considering the chevron-shaped structure in the
\Hbeta \ and \Hgamma \ emission lines, peaks at a positive delay and
is lower at $\tau = 0$ but does not extend to 0. This has to do with
the way MEMECHO deals with its delay map models. The default is set to
``pull'' the response down to zero at the ends of the delay map when
$\tau_{\rm min}$ and $\tau_{\rm max}$ are not equal to
zero. Therefore, when $\tau_{\rm min} = 0$, the default cannot pull
down at the low-$\tau$ end, so the lowest-entropy map is an
exponential function of $\tau$.  With $\tau_{\rm min} < 0$, the
entropy pulls down on both ends of the delay map, which then creates a
Gaussian peak at positive delays.

Because of this, different values of $\tau_{\rm min}$ can affect the
final velocity-delay maps produced, as this effectively regulates the
behavior of the model on some level when trying to smooth the delay
maps to the default image. We currently have no means to evaluate which
value of $\tau_{\rm min}$ results in the ``correct'' velocity-delay
map --- it is therefore worth considering velocity-delay maps with
various $\tau_{\rm min}$ parameters, as we do here, when drawing
conclusions about the BLR structure signatures, and any detailed
modeling and interpretations should consider and evaluate the
differences in these maps. In this work, we make only qualitative
statements regarding the signatures seen in the velocity-delay
maps. Whether or not the response actually reaches zero at any point
in the 3C\,120 maps is unknown, but for our purposes it is reassuring
that all of the maps show the same basic structures and kinematic
signatures. From Figure \ref{fig:ctmosaic}, comparing results for
$\tau_{\rm min} = -10, -1,$ and 0, we can also evaluate the evidence
for a deficit of prompt response from the resulting delay maps created
with each value of $\tau_{\rm min}$. At all three values of $\tau_{\rm
min}$, we see Gaussian delay distributions at positive lags in the
core of \Hbeta \ and \Hgamma \ in 3C\,120, but not for \heii \ ---
thus the data provide evidence for a deficit of prompt response in the
center of \Hbeta \ and \Hgamma, but not for \heii. This behavior is
consistent with our expectations for the stratification of the
BLR. 

\section{SUMMARY AND CONCLUSIONS}
\label{sec:summary}
We have presented velocity-delay maps constructed from the line and
continuum variations observed in four objects from our 2010
reverberation campaign. These maps provide new insights into the
structure of the BLR and constitute a dramatic increase in the number
of objects that have at least some information on the velocity field
of the BLR. Along with the velocity-delay maps for Arp 151
(\citealt{Bentz10b}) and the models of Mrk 50 (\citealt{Pancoast12}),
these velocity-delay maps provide the strongest constraints on the
structure of the BLR. Our velocity-delay map for 3C\,120 shows very
similar structure to the map of Arp 151, which also shows a lack of
prompt response in the Balmer lines. The continuum light curves for
Arp 151 were well-enough sampled that \cite{Bentz10b} were able to
recover a velocity-delay map using the original continuum light curve
instead of substituting in simulated ones. The similarities between
the maps for Arp 151 and our maps lends weight to the reliability of
our results using the simulated continuum light curves. We also see
very asymmetric profiles in both the Balmer and high-ionization
emission that is suggestive of infalling gas in all of our objects. In
3C\,120 and Mrk 335, the transfer function structure for the Balmer
lines differs from that for the \HeII \ emission line, suggesting
different structures dominating at different BLR radii. In all cases
where our data are of sufficient quality to constrain the structure of
the BLR, we see clear evidence of infall and rotation, both of which
result from the gravitational influence of the black hole. As
gravitationally dominated motion is the key assumption of
reverberation mapping, our results strongly support the reliability of
black hole mass estimates derived from reverberation mapping. Detailed
modeling of our most well-defined velocity-delay maps and a complete
discussion of the implications will follow in a future work.

\acknowledgments The authors are grateful for the support and
hospitality of the Dark Cosmology Centre for sponsorship of a workshop
where some of the critical phases of this work were carried out. BMP
and CJG gratefully acknowledge the support of the National Science
Foundation through grant AST-1008882 to The Ohio State University. KH
is supported by a Royal Society Leverhulme Trust Senior Research
Fellowship. BJS, CBH, and JLV are supported by NSF Fellowships. CSK,
AMM, and DMS acknowledge the support of NSF grants AST-1004756 and
AST-1009756. SK is supported at the Technion by the Kitzman Fellowship
and by a grant from the Israel-Niedersachsen collaboration program. SR
is supported at Technion by the Zeff Fellowship. SGS acknowledges the
support to CrAO in the frame of the `CosmoMicroPhysics' Target
Scientific Research Complex Programme of the National Academy of
Sciences of Ukraine (2007-2012). VTD acknowledges the support of the
Russian Foundation of Research (RFBR, project no. 12-02-01237-a). The
CrAO CCD cameras were purchased through the US Civilian Research and
Development for Independent States of the Former Soviet Union (CRDF)
awards UP1-2116 and UP1-2549-CR-03.


\begin{thebibliography}{47}
\expandafter\ifx\csname natexlab\endcsname\relax\def\natexlab#1{#1}\fi

\bibitem[{{Antokhin} \& {Bochkarev}(1983)}]{Antokhin83}
{Antokhin}, I.~I., \& {Bochkarev}, N.~G. 1983, \azh, 60, 448

\bibitem[{{Bahcall} {et~al.}(1972){Bahcall}, {Kozlovsky}, \&
  {Salpeter}}]{Bahcall72}
{Bahcall}, J.~N., {Kozlovsky}, B.-Z., \& {Salpeter}, E.~E. 1972, \apj, 171, 467

\bibitem[{{Bentz} {et~al.}(2009){Bentz}, {Peterson}, {Netzer}, {Pogge}, \&
  {Vestergaard}}]{Bentz09a}
{Bentz}, M.~C., {Peterson}, B.~M., {Netzer}, H., {Pogge}, R.~W., \&
  {Vestergaard}, M. 2009, \apj, 697, 160

\bibitem[{{Bentz} {et~al.}(2010{\natexlab{a}}){Bentz}, {Walsh}, {Barth},
  {Yoshii}, {Woo}, {Wang}, {Treu}, {Thornton}, {Street}, {Steele}, {Silverman},
  {Serduke}, {Sakata}, {Minezaki}, {Malkan}, {Li}, {Lee}, {Hiner}, {Hidas},
  {Greene}, {Gates}, {Ganeshalingam}, {Filippenko}, {Canalizo}, {Bennert}, \&
  {Baliber}}]{Bentz10a}
{Bentz}, M.~C., {et~al.} 2010{\natexlab{a}}, \apj, 716, 993

\bibitem[{{Bentz} {et~al.}(2010{\natexlab{b}}){Bentz}, {Horne}, {Barth},
  {Bennert}, {Canalizo}, {Filippenko}, {Gates}, {Malkan}, {Minezaki}, {Treu},
  {Woo}, \& {Walsh}}]{Bentz10b}
---. 2010{\natexlab{b}}, \apjl, 720, L46

\bibitem[{{Blandford} \& {McKee}(1982)}]{Blandford82}
{Blandford}, R.~D., \& {McKee}, C.~F. 1982, \apj, 255, 419

\bibitem[{{Bochkarev} \& {Antokhin}(1982)}]{Bochkarev82}
{Bochkarev}, N.~G., \& {Antokhin}, I.~I. 1982, Astronomicheskij Tsirkulyar,
  1238, 1

\bibitem[{{Capriotti} {et~al.}(1982){Capriotti}, {Foltz}, \&
  {Peterson}}]{Capriotti82}
{Capriotti}, E.~R., {Foltz}, C.~B., \& {Peterson}, B.~M. 1982, \apj, 261, 35

\bibitem[{{Clavel} {et~al.}(1989){Clavel}, {Wamsteker}, \& {Glass}}]{Clavel89}
{Clavel}, J., {Wamsteker}, W., \& {Glass}, I.~S. 1989, \apj, 337, 236

\bibitem[{{Clavel} {et~al.}(1990){Clavel}, {Boksenberg}, {Bromage}, {Elvius},
  {Penston}, {Perola}, {Santos-Lleo}, {Snijders}, \& {Ulrich}}]{Clavel90}
{Clavel}, J., {et~al.} 1990, \mnras, 246, 668

\bibitem[{{Clavel} {et~al.}(1991)}]{Clavel91}
---. 1991, \apj, 366, 64

\bibitem[{{Clavel}(1991)}]{Clavel91b} {Clavel}, J. 1991, {Variability
of Active Galaxies}, ed. C.~M. Duschl, S.~J. Wagner, \& M. Camenzind
(Berlin: Springer-Verlag), 31


\bibitem[{{Crenshaw} \& {Blackwell}(1990)}]{Crenshaw90}
{Crenshaw}, D.~M., \& {Blackwell}, Jr., J.~H. 1990, \apjl, 358, L37

\bibitem[{{Denney} {et~al.}(2011){Denney}, {Assef}, {Bentz}, {Dietrich},
  {Horne}, {Kochanek}, {Mathur}, {Peterson}, {Pogge}, \&
  {Vestergaard}}]{Denney11}
{Denney}, K., {et~al.} 2011, in Narrow-Line Seyfert 1 Galaxies and their Place
  in the Universe, Proceedings of Science, POS(NLS1)032

\bibitem[{{Denney} {et~al.}(2009){Denney}, {Peterson}, {Pogge}, {Adair},
  {Atlee}, {Au-Yong}, {Bentz}, {Bird}, {Brokofsky}, {Chisholm}, {Comins},
  {Dietrich}, {Doroshenko}, {Eastman}, {Efimov}, {Ewald}, {Ferbey}, {Gaskell},
  {Hedrick}, {Jackson}, {Klimanov}, {Klimek}, {Kruse}, {Lad{\'e}route}, {Lamb},
  {Leighly}, {Minezaki}, {Nazarov}, {Onken}, {Petersen}, {Peterson},
  {Poindexter}, {Sakata}, {Schlesinger}, {Sergeev}, {Skolski}, {Stieglitz},
  {Tobin}, {Unterborn}, {Vestergaard}, {Watkins}, {Watson}, \&
  {Yoshii}}]{Denney09c}
{Denney}, K.~D., {et~al.} 2009, \apjl, 704, L80

\bibitem[{{Denney} {et~al.}(2010){Denney}, {Peterson}, {Pogge}, {Adair},
  {Atlee}, {Au-Yong}, {Bentz}, {Bird}, {Brokofsky}, {Chisholm}, {Comins},
  {Dietrich}, {Doroshenko}, {Eastman}, {Efimov}, {Ewald}, {Ferbey}, {Gaskell},
  {Hedrick}, {Jackson}, {Klimanov}, {Klimek}, {Kruse}, {Lad{\'e}route}, {Lamb},
  {Leighly}, {Minezaki}, {Nazarov}, {Onken}, {Petersen}, {Peterson},
  {Poindexter}, {Sakata}, {Schlesinger}, {Sergeev}, {Skolski}, {Stieglitz},
  {Tobin}, {Unterborn}, {Vestergaard}, {Watkins}, {Watson}, \&
  {Yoshii}}]{Denney10}
---. 2010, \apj, 721, 715

\bibitem[{{Dietrich} {et~al.}(1993){Dietrich}, {Kollatschny}, {Peterson},
  {Bechtold}, {Bertram}, {Bochkarev}, {Boroson}, {Carone}, {Elvis},
  {Filippenko}, {Gaskell}, {Huchra}, {Hutchings}, {Koratkar}, {Korista},
  {Lame}, {Laor}, {MacAlpine}, {Malkan}, {Mendes de Oliveira}, {Netzer},
  {Penfold}, {Penston}, {Perez}, {Pogge}, {Richmond}, {Rosenblatt},
  {Shapovalova}, {Shields}, {Smith}, {Smith}, {Sun}, {Thiele}, {Veilleux},
  {Wagner}, {Wilkes}, {Wills}, \& {Wills}}]{Dietrich93}
{Dietrich}, M., {et~al.} 1993, \apj, 408, 416

\bibitem[{{Doroshenko} {et~al.}(2012){Doroshenko}, {Sergeev}, {Klimanov},
  {Pronik}, \& {Efimov}}]{Doroshenko12}
{Doroshenko}, V.~T., {Sergeev}, S.~G., {Klimanov}, S.~A., {Pronik}, V.~I., \&
  {Efimov}, Y.~S. 2012, \mnras, 426, 416

\bibitem[{{Ferland} {et~al.}(1992){Ferland}, {Peterson}, {Horne}, {Welsh}, \&
  {Nahar}}]{Ferland92}
{Ferland}, G.~J., {Peterson}, B.~M., {Horne}, K., {Welsh}, W.~F., \& {Nahar},
  S.~N. 1992, \apj, 387, 95

\bibitem[{{Gaskell}(1988)}]{Gaskell88}
{Gaskell}, C.~M. 1988, \apj, 325, 114

\bibitem[{{Grier} {et~al.}(2008)}]{Grier08}
{Grier}, C.~J., {et~al.} 2008, \apj, 688, 837

\bibitem[{{Grier} {et~al.}(2012){Grier}, {Peterson}, {Pogge}, {Denney},
  {Bentz}, {Martini}, {Sergeev}, {Kaspi}, {Minezaki}, {Zu}, {Kochanek},
  {Siverd}, {Shappee}, {Stanek}, {Araya Salvo}, {Beatty}, {Bird}, {Bord},
  {Borman}, {Che}, {Chen}, {Cohen}, {Dietrich}, {Doroshenko}, {Drake},
  {Efimov}, {Free}, {Ginsburg}, {Henderson}, {King}, {Koshida}, {Mogren},
  {Molina}, {Mosquera}, {Nazarov}, {Okhmat}, {Pejcha}, {Rafter}, {Shields},
  {Skowron}, {Szczygiel}, {Valluri}, \& {van Saders}}]{Grier12b}
---. 2012, \apj, 755, 60

\bibitem[{{Guerras} {et~al.}(2012){Guerras}, {Mediavilla}, {Jimenez-Vicente},
  {Kochanek}, {Mu{\~n}oz}, {Falco}, \& {Motta}}]{Guerras12}
{Guerras}, E., {Mediavilla}, E., {Jimenez-Vicente}, J., {Kochanek}, C.~S.,
  {Mu{\~n}oz}, J.~A., {Falco}, E., \& {Motta}, V. 2012, arXiv:1207:2042

\bibitem[{{Horne}(1994)}]{Horne94}
{Horne}, K. 1994, in Astronomical Society of the Pacific Conference Series,
  Vol.~69, Reverberation Mapping of the Broad-Line Region in Active Galactic
  Nuclei, ed. P.~M. {Gondhalekar}, K.~{Horne}, \& B.~M. {Peterson}, 23--25

\bibitem[{{Horne} {et~al.}(2004){Horne}, {Peterson}, {Collier}, \&
  {Netzer}}]{Horne04}
{Horne}, K., {Peterson}, B.~M., {Collier}, S.~J., \& {Netzer}, H. 2004, \pasp,
  116, 465

\bibitem[{{Horne} {et~al.}(1991){Horne}, {Welsh}, \& {Peterson}}]{Horne91}
{Horne}, K., {Welsh}, W.~F., \& {Peterson}, B.~M. 1991, \apjl, 367, L5

\bibitem[{{Kaspi} {et~al.}(2000){Kaspi}, {Smith}, {Netzer}, {Maoz}, {Jannuzi},
  \& {Giveon}}]{Kaspi00}
{Kaspi}, S., {Smith}, P.~S., {Netzer}, H., {Maoz}, D., {Jannuzi}, B.~T., \&
  {Giveon}, U. 2000, \apj, 533, 631

\bibitem[{{Kelly} {et~al.}(2009){Kelly}, {Bechtold}, \&
  {Siemiginowska}}]{Kelly09}
{Kelly}, B.~C., {Bechtold}, J., \& {Siemiginowska}, A. 2009, \apj, 698, 895

\bibitem[{{Kollatschny}(2003)}]{Kollatschny03}
{Kollatschny}, W. 2003, \aap, 407, 461

\bibitem[{{Kollatschny} \& {Dietrich}(1997)}]{Kollatschny97}
{Kollatschny}, W., \& {Dietrich}, M. 1997, \aap, 323, 5

\bibitem[{{Koratkar} \& {Gaskell}(1989)}]{Koratkar89}
{Koratkar}, A.~P., \& {Gaskell}, C.~M. 1989, \apj, 345, 637

\bibitem[{{Koz{\l}owski} {et~al.}(2010){Koz{\l}owski}, {Kochanek}, {Udalski},
  {Wyrzykowski}, {Soszy{\'n}ski}, {Szyma{\'n}ski}, {Kubiak}, {Pietrzy{\'n}ski},
  {Szewczyk}, {Ulaczyk}, {Poleski}, \& {The OGLE Collaboration}}]{Kozlowski10}
{Koz{\l}owski}, S., {et~al.} 2010, \apj, 708, 927

\bibitem[{{Krolik} {et~al.}(1991){Krolik}, {Horne}, {Kallman}, {Malkan},
  {Edelson}, \& {Kriss}}]{Krolik91}
{Krolik}, J.~H., {Horne}, K., {Kallman}, T.~R., {Malkan}, M.~A., {Edelson},
  R.~A., \& {Kriss}, G.~A. 1991, \apj, 371, 541

\bibitem[{{MacLeod} {et~al.}(2010){MacLeod}, {Ivezi{\'c}}, {Kochanek},
  {Koz{\l}owski}, {Kelly}, {Bullock}, {Kimball}, {Sesar}, {Westman}, {Brooks},
  {Gibson}, {Becker}, \& {de Vries}}]{MacLeod10}
{MacLeod}, C.~L., {et~al.} 2010, \apj, 721, 1014

\bibitem[{{MacLeod} {et~al.}(2012){MacLeod}, {Ivezi{\'c}}, {Sesar}, {de Vries},
  {Kochanek}, {Kelly}, {Becker}, {Lupton}, {Hall}, {Richards}, {Anderson}, \&
  {Schneider}}]{MacLeod12}
---. 2012, \apj, 753, 106

\bibitem[{{Pancoast} {et~al.}(2011){Pancoast}, {Brewer}, \&
  {Treu}}]{Pancoast11}
{Pancoast}, A., {Brewer}, B.~J., \& {Treu}, T. 2011, \apj, 730, 139

\bibitem[{{Pancoast} {et~al.}(2012){Pancoast}, {Brewer}, {Treu}, {Barth},
  {Bennert}, {Canalizo}, {Filippenko}, {Gates}, {Greene}, {Li}, {Malkan},
  {Sand}, {Stern}, {Woo}, {Assef}, {Bae}, {Buehler}, {Cenko}, {Clubb},
  {Cooper}, {Diamond-Stanic}, {Hiner}, {H{\"o}nig}, {Joner}, {Kandrashoff},
  {Laney}, {Lazarova}, {Nierenberg}, {Park}, {Silverman}, {Son}, {Sonnenfeld},
  {Thorman}, {Tollerud}, {Walsh}, \& {Walters}}]{Pancoast12}
{Pancoast}, A., {et~al.} 2012, \apj, 754, 49

\bibitem[{{Peterson}(1993)}]{Peterson93}
{Peterson}, B.~M. 1993, \pasp, 105, 247

\bibitem[{{Peterson} {et~al.}(1991)}]{Peterson91}
{Peterson}, B.~M., {et~al.} 1991, \apj, 368, 119

\bibitem[{{Peterson} {et~al.}(2004)}]{Peterson04}
---. 2004, \apj, 613, 682

\bibitem[{{Sergeev} {et~al.}(2007){Sergeev}, {Doroshenko}, {Dzyuba},
  {Peterson}, {Pogge}, \& {Pronik}}]{Sergeev07}
{Sergeev}, S.~G., {Doroshenko}, V.~T., {Dzyuba}, S.~A., {Peterson}, B.~M.,
  {Pogge}, R.~W., \& {Pronik}, V.~I. 2007, \apj, 668, 708

\bibitem[{{Ulrich} \& {Horne}(1996)}]{Ulrich96}
{Ulrich}, M.-H., \& {Horne}, K. 1996, \mnras, 283, 748

\bibitem[{{van Groningen} \& {Wanders}(1992)}]{vanGroningen92}
{van Groningen}, E., \& {Wanders}, I. 1992, \pasp, 104, 700

\bibitem[{{Wanders} \& {Peterson}(1996)}]{Wanders96}
{Wanders}, I., \& {Peterson}, B.~M. 1996, \apj, 466, 174

\bibitem[{{Wanders} {et~al.}(1995){Wanders}, {Goad}, {Korista}, {Peterson},
  {Horne}, {Ferland}, {Koratkar}, {Pogge}, \& {Shields}}]{Wanders95}
{Wanders}, I., {et~al.} 1995, \apjl, 453, L87

\bibitem[{{Zu} {et~al.}(2012){Zu}, {Kochanek}, {Koz{\l}owski}, \&
  {Udalski}}]{Zu12}
{Zu}, Y., {Kochanek}, C.~S., {Koz{\l}owski}, S., \& {Udalski}, A. 2012,
arXiv:1207.3794

\bibitem[{{Zu} {et~al.}(2011){Zu}, {Kochanek}, \& {Peterson}}]{Zu11}
{Zu}, Y., {Kochanek}, C.~S., \& {Peterson}, B.~M. 2011, \apj, 735, 80

\end{thebibliography}

\clearpage

\begin{deluxetable}{lccc}
\tablewidth{0pt}
\tablecaption{Object List}
\tablehead{
\colhead{ } &
\colhead{ } &
\colhead{ } &
\colhead{ } \\
\colhead{Object} &
\colhead{RA} &
\colhead{DEC} &
\colhead{$z$}  \\
\colhead{} &
\colhead{(J2000)} &
\colhead{(J2000)} &
\colhead{} 
} 
\startdata
Mrk 335      & 00 06 19.5 & +20 12 10 & 0.0258 \\
Mrk 1501     & 00 10 31.0 & +10 58 30 & 0.0893 \\
3C\,120      & 04 33 11.1 & +05 21 16 & 0.0330 \\
Mrk 6        & 06 52 12.2 & +74 25 37 & 0.0188 \\
PG\,2130+099 & 21 32 27.8 & +10 08 19 & 0.0630  	
\enddata
\label{Table:obj_info}
\end{deluxetable} 


\begin{deluxetable}{lcccc}
\tablewidth{0pt}
\tablecaption{MEMECHO Parameters\tablenotemark{a} }
\tablehead{
\colhead{ } &
\colhead{ } &
\colhead{ } &
\colhead{ } &
\colhead{ } \\
\colhead{Object} &
\colhead{$A$} &
\colhead{$W$} &
\colhead{$\tau_{\rm min}$} &
\colhead{$\chi^2/N$}  
} 
\startdata
Mrk 335      & 0.1 & 1.0 & -1 & 2.8 \\
Mrk 1501     & 1.0 & 1.0 & 0  & 1.8 \\
3C\,120      & 1.0 & 1.0 & 0  & 2.4 \\
PG\,2130+099 & 0.1 & 1.0 & 0  & 2.1 	
\enddata
\label{Table:MEMECHO}
\tablenotetext{a}{As defined in Section \ref{sec:memecho}.}
\end{deluxetable}

\clearpage




\end{document}